\documentclass[a4paper, 10pt, twocolumn, notitlepage]{article}

\usepackage{amssymb, amsfonts, amsmath, eucal, tikz, pgfplots, graphicx,
subfigure, array, fancyhdr,multicol, multirow, pgfplots, setspace, titlesec,
titletoc,titling}
\usepackage[runin]{abstract}

\usetikzlibrary{shapes,patterns,arrows,svg.path}
\pgfplotsset{compat=1.3}
 
\usepackage{bm}
\usepackage{graphics, color, mathrsfs, authblk}
\usepackage{psfrag}
\usepackage{MnSymbol}

\newcommand{\degrees}{\ensuremath{^\circ}}
\definecolor{red}{rgb}{1.0,0.0,0.0}
\definecolor{blue}{rgb}{0.0,0.0,1.0}
\definecolor{custom1}{rgb}{0,0,0.2353}
\definecolor{custom2}{rgb}{0,0,0.9882}
\definecolor{custom3}{rgb}{0.7751,0.1561,1}
\definecolor{custom4}{rgb}{1,0.6580,0.3420}
\definecolor{custom5}{rgb}{1,1,0.2157}

\definecolor{rulecolor}{rgb}{0.6,0.6,0.6}
\newcommand{\HRule}{\textcolor{rulecolor}{\rule[0.25cm]{\linewidth}{0.3mm}}}  

\titleformat{\section}{\large\sffamily\bfseries}{\thesection\quad}{0pt}{}[\HRule]
\titlespacing*{\section}{0cm}{0.3cm}{-0.1cm}

\titleformat{\subsection}{\normalsize\sffamily\bfseries}{\thesubsection\quad}{0pt}{}
\titlespacing*{\subsection}{0pt}{0.3cm}{0.1cm}

\pretitle{\begin{flushleft}\LARGE\sffamily\bfseries}
\posttitle{\par\end{flushleft}\vskip 0.4em}
\preauthor{\begin{flushleft}\normalsize\rmfamily}
\postauthor{\par\end{flushleft}\vskip 0.1em}
\predate{\begin{flushleft}\small\sc\rmfamily}
\postdate{\par\end{flushleft}\vskip 0.1em}

\title{Modelling the Fluid Mechanics of Cilia and Flagella in Reproduction and Development}
\author[1,2]{Thomas D. Montenegro-Johnson}
\author[1,2]{Andrew A. Smith}
\author[1,2,3]{David J. Smith}
\author[1]{Daniel Loghin}
\author[1,2]{John R. Blake}
\affil[1]{School of Mathematics, University of Birmingham, Edgbaston,
Birmingham, B15 2TT, UK}
\affil[2]{Centre for Human Reproductive Science, Birmingham Women's NHS
Foundation Trust, Edgbaston, Birmingham, B15 2TG, UK}
\affil[3]{{School of Engineering \& Centre for Scientific Computing, University
of Warwick, Coventry, CV4 7AL, UK}}

\date{}

\begin{document}


\newlength\fheight
\newlength\fwidth

\twocolumn[
  \begin{@twocolumnfalse}
    \maketitle
    \vspace{-1.7cm}
    \begin{abstract}
    Cilia and flagella are actively bending slender organelles, performing functions
such as motility, feeding and embryonic symmetry breaking.  We review the
mechanics of viscous-dominated microscale flow, including time-reversal
symmetry, drag anisotropy of slender bodies, and wall effects.  We focus on the
fundamental force singularity, higher order multipoles, and the method of
images, providing physical insight and forming a basis for computational
approaches. Two biological problems are then considered in more detail: (1)
left-right symmetry breaking flow in the node, a microscopic structure in
developing vertebrate embryos, and (2) motility of microswimmers through
non-Newtonian fluids. Our model of the embryonic node reveals how particle
transport associated with morphogenesis is modulated by the gradual emergence of
cilium posterior tilt.  Our model of swimming makes use of force distributions
within a body-conforming finite element framework, allowing the solution of
nonlinear inertialess Carreau flow.  We find that a three-sphere model swimmer
and a model sperm are similarly affected by shear-thinning; in both cases
swimming due to a prescribed beat is enhanced by shear-thinning, with optimal
Deborah number around 0.8.  The sperm exhibits an almost perfect linear
relationship between velocity and the logarithm of the ratio of zero to infinite
shear viscosity, with shear-thickening hindering cell progress.
Montenegro-Johnson et al, Eur.\ Phys.\ J.\ E, 35 10 (2012) 111, published online: 29 October 2012 
doi: 10.1140/epje/i2012-12111-1.  \bigskip\bigskip
    \end{abstract}
  \end{@twocolumnfalse}
]


\section{Introduction}
\label{sec:introduction}
The active locomotion of cells and transport of fluids on microscopic scales has
been a benchmark problem in applied mathematics for the past 60 years, since
Taylor \cite{Taylor51} demonstrated that a two-dimensional sheet could swim
utilising only viscous forces. The field had been an active area of research for
zoologists for some time
\cite{engelmann1868flimmerbewegung,verworn1891studien,parker1905movements,gray1928ciliary},
though a leap forward in understanding was made when experimentalists and
theoreticians began to collaborate. It was in this spirit of collaboration,
fostered by Taylor and Gray, that Hancock \cite{Hancock53} first developed
slender body theory, a powerful method based upon modelling slender swimmers
{by distributions} of force singularities, which in turn has led
to the development of the singularity methods used in the present study.

Microscale fluid propulsion is usually achieved in nature through the beating of
cilia and flagella. These are slender, hair-like organelles that perform a range
of functions from locomotion to sensory reception. In reproduction, flagella
propel sperm cells, allowing sperm and egg to meet, and cilia transfer the
fertilised embryo from the ampulla to the uterus.  Then, in the early stages of
{vertebrate} embryonic development, cilia are responsible for
the production of a directional fluid flow which breaks left-right symmetry in
vertebrates \cite{Nonaka98}. This occurs in a fluid filled cavity that appears
on the embryo shortly after fertilisation, called the node.

\begin{figure}[t!]
	\centering
	\includegraphics*[scale=0.4, viewport=0 150 600 700,clip]{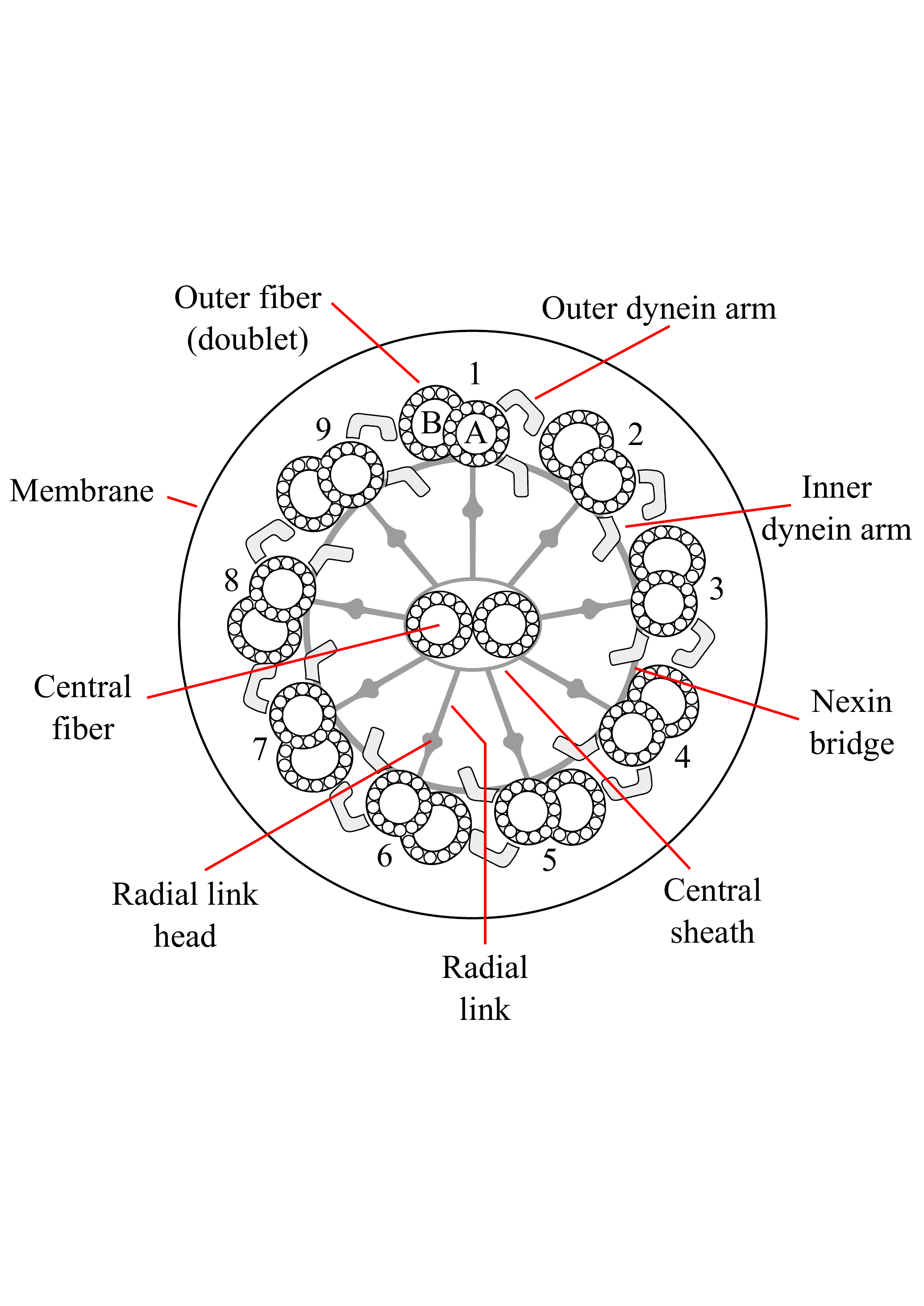}
	\caption{A schematic cross-section of the ``9+2'' axoneme, redrawn from
	Fawcett \cite{fawcett1975mammalian}.}
	\label{fig:axoneme}
\end{figure}

In eukaryotic cells, cilia and flagella induce active bending along their length
via a remarkable, evolutionarily-conserved internal structure known as the
\textit{axoneme}, which was discovered with the advent of electron microscopy
\cite{manton1952electron,fawcett1954,afzelius1959}. The axoneme comprises 9
inextensible outer microtubule doublets, as shown in fig.~\ref{fig:axoneme}, and
passive linking elements which stiffen the assembly. The combination of
relative, localised microtubule sliding, their inextensibility and the
restraining effects of linking structures, generates bending. This is the
`sliding filament theory' first demonstrated by Satir \cite{satir1965studies}.
For human sperm, as with most motile cilia and flagella, a central pair of
microtubules runs along the length of the axoneme.  This configuration is
referred to as the ``9+2'' axoneme. Nodal cilia, however, lack this central
pair. These ``9+0'' cilia were thought to be immotile until the relatively
recent work of Nonaka \textit{et al.} \cite{Nonaka98}, which showed that they
`whirled' with a near rigid-body motion, quite distinct from the extensively
studied beat patterns of 9+2 cilia.

In Newtonian fluids, such as sea water, microscopic cilia and flagella must
execute a beat pattern that is not time reversible \cite{purcell1977life} in
order to generate a net fluid flow. This is due to the lack of time dependence
in the governing fluid equations, which are discussed in
sect.~\ref{sec:fluid_dynamics}. For sperm flagella, time reversibility is broken
by propagating a bending wave along the length of the flagellum. However, nodal
cilia are incapable of executing sophisticated waveforms, and so instead break
symmetry by tilting their axis of rotation in a given direction. Hydrodynamic
interaction with the cell wall then leads to an effective-recovery stroke
asymmetry.

In this paper, we first summarise the fluid mechanics that govern the flow
generated by cilia and flagella. We discuss the flow solution generated by a
singular force, and the relationship between this fundamental solution and the
physics governing the flow induced by the motion of slender bodies. We show how
the effects of a no-slip wall may be incorporated through the inclusion of
`image systems' {involving} higher-order flow singularities,
such as point stresses and torques. These are demonstrated graphically via
singularity diagrams. We then discuss computational techniques that have arisen
from modelling the action of cilia and flagella on the fluid by distributions of
these flow singularities.

A combination of these computational techniques is used to model cilia-driven
flow in the embryonic node in mice (fig.~\ref{fig:mouse_node}). The
node is modelled at various stages of development, and the effects of posterior
cilium tilt on the generation of directional flow is examined.

For many biological flow problems, for instance the swimming of sperm through
cervical mucus, the assumption of Newtonian fluid rheology, as discussed in
sect.~\ref{sec:fluid_dynamics}, is invalid. In such cases, the fluid dynamics is
governed by complicated nonlinear equations. We apply a {finite
element} method developed by {Montenegro-Johnson} \textit{et al.}
\cite{johnson2012femlets} to an artificial swimmer described by Najafi and
Golestanian \cite{najafi2004simple}, and examine the effects of nonlinear fluid
rheology on the swimmer's progression. We then apply the same techniques to
model a two-dimensional analogue of human sperm.

\begin{figure}[tbp]
	\centering
	\caption{Copyright figure available in published version. (a) Electron
micrograph of a mouse embryo at 7.5 days post fertilisation, indicating the
anterior-posterior and left-right directions {(VN, ventral node; NP, notochordal
plate; FG, foregut)}. Bar $100\,\mu$m, reprinted from Hirokawa \textit{et al.\
}\cite{Hirokawa06} (b)
Cilia covering the nodal pit cells{, reprinted} from Nonaka \textit{et al.\
}\cite{Nonaka05}. (c) Schematic figure showing a simplified model for nodal
flow. Cilia rotate, with axis of rotation tilted towards the posterior, creating
a right-to-left flow above the cilia tips. This is balanced by a return flow,
due to the overlying Reichert's membrane. The flow is believed to transport
Nodal Vesicular Parcels (NVPs) which break up at the left of the node,
delivering morphogen proteins to initiate asymmetric development.}
	\label{fig:mouse_node}
\end{figure}



\section{Fluid mechanics of cilia and flagella}
\label{sec:fluid_dynamics}
By considering the forces acting on an arbitrary volume of fluid and applying
the laws of conservation of momentum and mass, we may derive the Cauchy equations
\begin{subequations}
	\begin{align}
	&\rho \left( \frac{\partial \bm{u}}{\partial t} + (\bm{u}\cdot \nabla)
	\bm{u}\right) = \nabla \cdot \boldsymbol{\sigma} + \bm{F},
	\label{eq:momentum_conservation} \\
	&\frac{\partial \rho}{\partial t} + \nabla \cdot (\rho \bm{u}) = 0,
	\end{align}
	\label{eq:mass_conservation}
\end{subequations}
which govern the motion of the fluid. Here, $\bm{F}$ {is} the body force
acting on the fluid, such as gravity, $\rho$ is the fluid density and
$\bm{u}$ is the fluid velocity in a fixed frame of reference. The stress
tensor $\boldsymbol{\sigma}$ incorporates the forces acting over the surface of
an arbitrary parcel of fluid, such as pressure and internal friction, and is
dependent on the type of fluid being modelled.

For Newtonian fluids, stress is proportional to strain {rate}, so that the
fluid viscosity depends only on temperature, which is assumed to be constant
throughout this study. In such cases,
\begin{equation}
	\boldsymbol{\sigma} = - p \bm{I} +
	\frac{\mu}{2}\boldsymbol{\varepsilon}(\bm{u}),
\label{eq:newtonian_stress}
\end{equation}
for pressure, $p$ and strain rate $\boldsymbol{\varepsilon}(\bm{u}) = \nabla
\bm{u} + (\nabla \bm{u})^T$. A non-dimensionalisation of the
Navier-Stokes equations that result from substitution of the stress
\eqref{eq:newtonian_stress} into the Cauchy momentum equation
\eqref{eq:momentum_conservation} shows that the relative importance of viscous
forces to inertial forces is given by the Reynolds number
\begin{equation}
\mathrm{Re} = \frac{\rho U^2 L^2}{\mu U L} = \frac{\mathrm{inertial\
force}}{\mathrm{viscous\ force}}.
\end{equation}

Fluid pumping and locomotion by microscopic cilia and flagella entails typical
length-scales $L$ of $\mathcal{O}(10^{-5} - 10^{-4})\,\mathrm{m}$ and velocities
$U$ of $\mathcal{O}(10^{-5} - 10^{-4})\,\mathrm{m}\cdot\mathrm{s}^{-1}$, with
typical fluid densities around $\rho =
\mathcal{O}(10^3)\,\mathrm{kg}\cdot\mathrm{m}^{-3}$ and fluid viscosities of
$\mu = \mathcal{O}(10^{-3})\,\mathrm{Pa}\cdot\mathrm{s}$ or greater. Thus, the
Reynolds number of the flows we will consider is no higher than $\mathrm{Re} =
10^{-2} \ll 1$, which indicates that viscous forces dominate over inertial
forces. Thus, for Newtonian fluids, an accurate representation of the fluid
mechanics is given by the Stokes flow equations,
\begin{equation}
	\mu\nabla^2\bm{u} -\nabla p + \bm{F} = 0, \quad
	\nabla\cdot\bm{u} = 0. \label{eq:stokes}
\end{equation}

\subsection{The fundamental singularity}

Insight into biological flows generated by cilia and flagella may be gained by
considering the flows that arise due to concentrated driving forces.  Consider
an infinite fluid obeying the Stokes {flow} equations \eqref{eq:stokes} that is driven
by a concentrated force per unit volume $\bm{F} = \bm{f}
\delta(\bm{x}-\bm{y})$, of magnitude and direction $\bm{f}$, where
$\delta$ is the Dirac delta distribution centered
at $\bm{y}$. The
velocity solution corresponding to this fundamental singularity is given by
\begin{equation}
	u_i(\bm{x}) = \frac{1}{8\pi\mu}\left(\frac{\delta_{ij}}{r} +
	\frac{r_ir_j}{r^3}\right)f_j(\bm{y}) = S_{ij}(\bm{x},\bm{y})f_j(\bm{y}),
	\label{eq:stokeslet}
\end{equation}
where $r_i = x_i - y_i$ for $i = 1,2,3$, $r^2 = r_1^2 + r_2^2 + r_3^2$ and
$S_{ij}(\bm{x},\bm{y})$ is known as the Stokeslet, or Oseen-Burgers tensor.
{The Einstein summation convention applies throughout.}

\begin{figure}[htbp]
	\centering
	\includegraphics{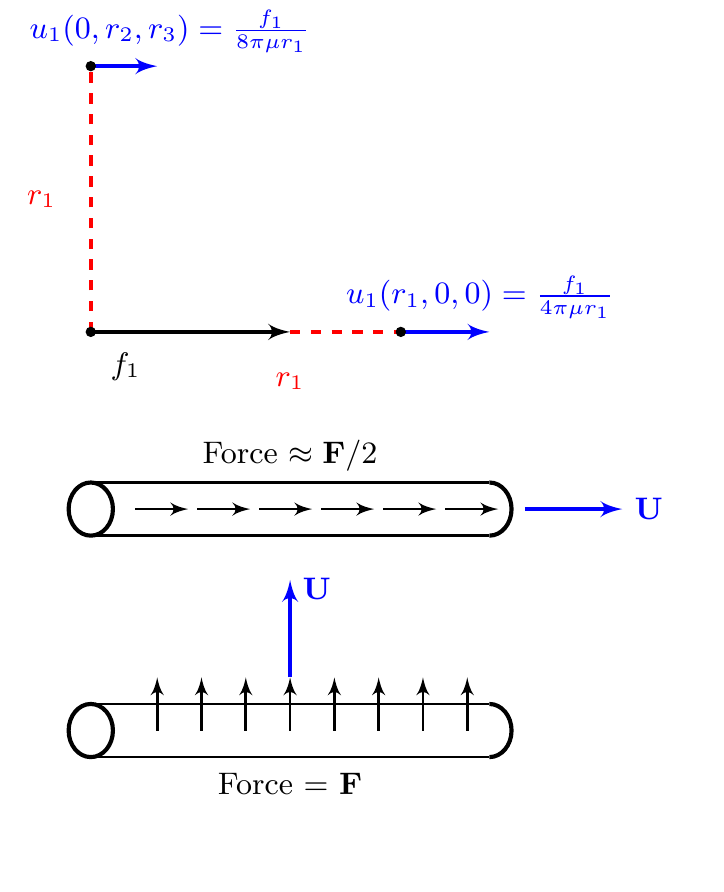}
	\caption{The flow generated by a singular force, and a demonstration of
	the 2:1 drag anisotropy that enables flagellar and ciliary propulsion,
	redrawn from Blake and Sleigh \cite{Blake74b}.}
	\label{fig:two_to_one}
\end{figure}

The anisotropy of the Stokeslet is an important factor affecting the fluid
mechanics of cilia and flagella. As illustrated in fig.~\ref{fig:two_to_one},
the velocity of the flow due to a singular force, at a distance $r_1$ from the
force, is twice as large at points in line with the force {than
at those points} perpendicular to the force
at the same distance. As a first approximation, the action of a slender cylinder
moving through a fluid may be represented by a line distribution of singular
driving forces, as shown in fig.~\ref{fig:two_to_one}. Since the Stokes flow
equations~\eqref{eq:stokes} are linear, the corresponding velocity field is
given by a sum of Stokeslet solutions. Thus, the drag on a slender body moving
tangentially through the flow is approximately half that on an equivalent body
moving normally. This ``two to one ratio'', first described in the resistive
force theory of Gray and Hancock \cite{Gray55} is the basis for flagella and
ciliary propulsive dynamics. However, drag anisotropy is not itself essential
for very low Reynolds number propulsion if the filaments are extensible, as
recently shown by Pak \& Lauga \cite{pak2011}.

\begin{figure}
\begin{center}
\includegraphics[scale=0.8]{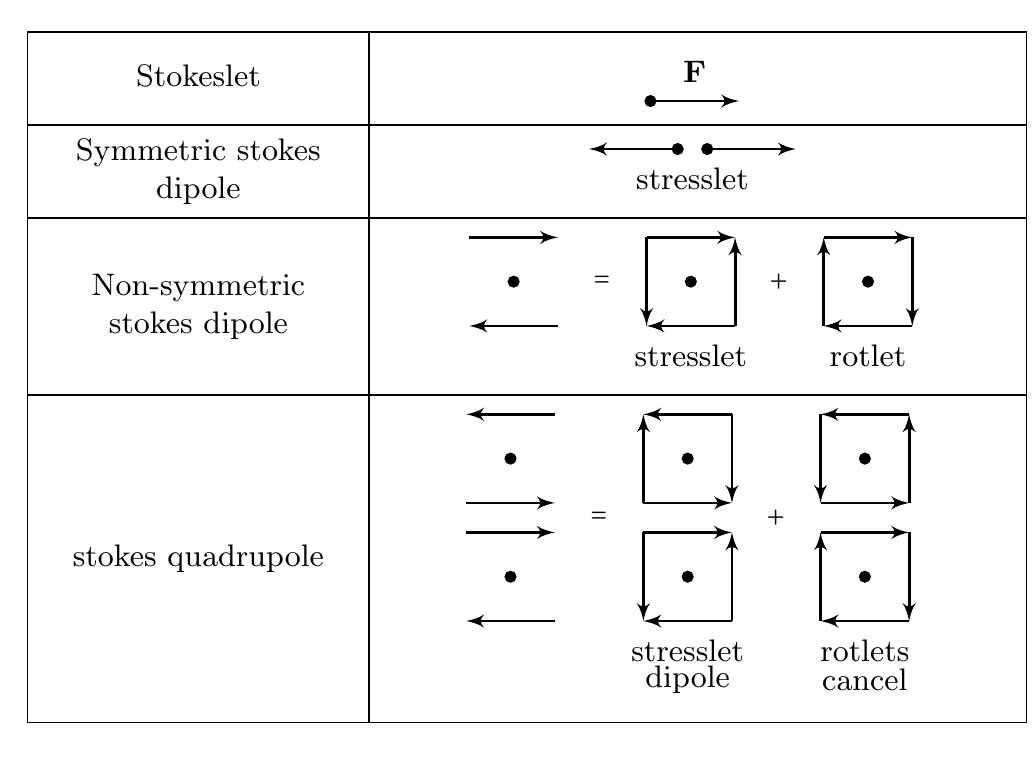}
\end{center}
\caption{Singularities of Stokes flow, with forces represented by arrow vectors.
Displaced forces represent higher order singularities.}
\label{fig:sing_diag}
\end{figure}

Swimmers in Stokes flow move in such a way that no net forces \cite{Taylor51} or
torques \cite{chwang1971note} act upon them, and flows driven by cilia
protruding from epithelia can be modelled with image systems, as discussed in
sect.~\ref{sec:wall}. In both cases, the far-fields of the flow are given by
higher-order singularities. By taking derivatives of the Stokeslet, it is
possible to derive the flow fields generated by higher-order singularities, such
as point stresses and point torques. These are able to provide valuable insight
into the far-field behaviour of the fluid surrounding swimming cells, and into
the hydrodynamic effects arising from the inclusion of no-slip boundaries in the
flow \cite{Blake71a,vilfan2006hydrodynamic}. Fig.~\ref{fig:sing_diag} shows the
schematic representation of some of these singularities.

With each increase in the order of singularity, the decay of the fluid velocity
in the far-field is increased by $\mathcal{O}(1/r)$, so that Stokeslets
decay with $\mathcal{O}(1/r)$, stokes dipoles decay with $\mathcal{O}(1/r^2)$
and stokes quadrupoles with $\mathcal{O}(1/r^3)$.

An additional important singularity, familiar as the source dipole of potential flow
theory, is given by,
\begin{equation}
	D_{ij}(\bm{x},\bm{y}) = -\frac{1}{4\pi}\left(\frac{\delta_{ij}}{r^3} -
	\frac{3r_ir_j}{r^5}\right).
	\label{eq:dipole}
\end{equation}
This expression for the $i$-component of a velocity field due to a source dipole oriented in the $j$-direction,
together with zero pressure field, is also a solution of the Stokes flow equations, and can be combined with the
Stokeslet to formulate solutions for translating spheres, ellipsoids and slender rods. The source dipole can alternatively
be identified as the Laplacian of the Stokeslet---in other words, a particular form of the stokes quadrupole.

\begin{figure}[htbp]
	\centering
    \includegraphics{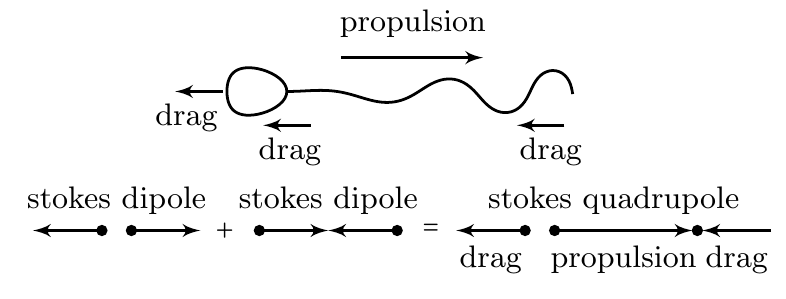}
	\caption{An approximate singularity representation of the flow field surrounding a human sperm,
	redrawn from Smith and Blake \cite{smith2009ms}. The quadrupole
	representation was suggested after calculation of the force
	distribution in the tail with slender body theory.}
	\label{fig:sperm_singularity}
\end{figure}

Singularity models capture many of the essential features of cilia and flagella
driven flows. Fig.~\ref{fig:experimental} shows experimental data from recent
studies by Drescher \textit{et al.\ }\cite{drescher2010,drescher2011fluid}.
Sperm, bacteria and individual algae are too small for gravitational
sedimentation to have a dominant effect on the flow field; the zero total force
condition therefore entails that the far-field is given by a stresslet
{(fig.~\ref{fig:experimental}a)}. Simulation modelling
(fig.~\ref{fig:sperm_singularity}) predicts that the flow field around a sperm
is approximated well by a stokes quadrupole, given by drag components at the
front and rear of the cell and a propulsive component in the middle;
sufficiently far from the cell the dominant singularity will however still be
the stresslet.

The flow field closer to the cell is more complex; for biflagellate algae the
time averaged flow field due to two propulsive {flagella} and
{the} cell body can be represented by a three Stokeslet model
\cite{drescher2010} {(fig.~\ref{fig:experimental}b)}. Larger
swimmers such as \textit{Volvox Carteri} colonies are subject to a significant
gravitational force, evident as a Stokeslet far-field
(fig.~\ref{fig:experimental}c); the near-field is given by a source dipole and
stresslet combination (fig.~\ref{fig:experimental}d).

\begin{figure}[htbp]
	\centering
		\caption{Copyright figure available in published version. (a) Experimentally observed stresslet behaviour in the
		far-field of a swimming bacterium \cite{drescher2011fluid}
		 (b) Experimentally observed flow field
		around a swimming biflagellate \textit{Chlamydomonas
		Reinhardtii}, showing that the data are fitted accurately by a three-Stokeslet model \cite{drescher2010} (c,d) Experimentally observed flow fields around a larger swimmer, the \textit{Volvox Carteri} algal colony {\cite{drescher2010}}. The larger swimmer is subject to a non-negligible gravitational force, evident as a far-field Stokeslet. (d) Shows the near-field when the Stokeslet field is subtracted, observed to resemble a source dipole and stresslet \cite{drescher2010} }
            \label{fig:experimental}
\end{figure}

\subsection{Wall effects} \label{sec:wall}

The presence of a rigid boundary near a Stokeslet significantly alters the
resultant flow field. This may be understood through the use of singularity
diagrams. To enforce the zero velocity condition on the plane boundary, that may
for example represent the nodal floor, an image system of singularities is
placed the other side of the boundary \cite{Blake71a}. The explicit form for the
velocity field arising from a point force located at height $h$ above a plane boundary represented by image systems is,
\begin{align}
B_{ij}(\bm{x},\bm{y}) = \frac{1}{8\pi\mu}& \left[\left(\frac{\delta_{ij}}{r} + \frac{r_ir_j}{r^3}\right) - \left(\frac{\delta_{ij}}{R} + \frac{R_iR_j}{R^3}\right) \right. \notag \\
&\hspace{-3em}  \left. + 2h \Delta_{jk}
\frac{\partial}{\partial R_k}\left\{\frac{hR_i}{R^3} - \left(\frac{\delta_{i3}}{R} + \frac{R_iR_3}{R^3}\right)\right\} \right] \mbox{.}
\label{eq:blakelet}
\end{align}
The tensor $\Delta_{jk}$ takes value $+1$ for $j=k=1,2$, value $-1$ for $j=k=3$, and zero if $j\neq k$. The image location is given by $R_1=r_1$,
$R_2=r_2$ and $R_3=-h$.

The {image systems} for the two cases corresponding to a force orientated parallel
and perpendicular to the boundary are shown in fig.~\ref{fig:wall_effects}.
The far-field arising from a Stokeslet near a plane boundary is of higher order
than the corresponding Stokeslet in an infinite fluid, being
$\mathcal{O}(1/r^2)$ in the parallel case and $\mathcal{O}(1/r^3)$ for the
perpendicular case. The boundary therefore has the effect of shielding the
fluid in the far-field from the effect of the Stokeslet. It is this shielding
effect, illustrated in fig.~\ref{fig:cilium_field}, that allows nodal cilia to
generate directional fluid flow by tilting their axis of rotation in the
posterior direction in mice \cite{Smith07}; we will investigate the resultant
fluid mechanics in more detail in sect.~\ref{nodal}.
\begin{figure}[htbp]
	\begin{center}
\end{center}
	\caption{Copyright figure available in published version. A diagram illustrating the image systems for a Stokeslet
	orientated parallel (a) and perpendicular (b) to a no-slip plane
	boundary. Note that the wall induces a more rapid decay in the
	far-field, which is of $\mathcal{O}(1/r^2)$ in the parallel case and
	$\mathcal{O}(1/r^3)$ in the perpendicular case. \cite[fig. 4(a,b)]{Smith11}
      }
	\label{fig:wall_effects}
\end{figure}
\begin{figure}[htbp]
	\centering


	\caption{The \textit{wall effect}: the zones of influence of a whirling nodal cilium protruding from a cell surface during (a) the
	effective stroke and (b) the recovery stroke. The far-field strength
	decays more rapidly than the inner- and near-fields due to the wall
	influence; moreover, during the effective stroke, the cilium is further
	from the cell surface than during the recovery stroke, resulting in the
	near-field having greater extent, and propelling more fluid. This
	effective-recovery stroke asymmetry results in net fluid propulsion in
	the direction of the effective stroke.}
	\label{fig:cilium_field}
\end{figure}

For further detailed review of low Reynolds number biofluiddynamics, see Dillon
\textit{et al.} \cite{dillon2007fluid}, Lauga and Powers
\cite{lauga2009hydrodynamics} and Gaffney \textit{et al.}
\cite{gaffney2011mammalian}.
We now consider two problems which can be analysed using singularity approaches.
For the first, Newtonian symmetry-breaking flow in the node, we apply slender
body theory and the method of regularised Stokeslets to model the cilia and the
node geometry respectively{, discussed in sect.~\ref{nodal}}. For the second, swimming in non-Newtonian fluids, we
use a recently developed hybrid of singularity methods with the finite element
method \cite{johnson2012femlets}, which we have dubbed the `method of femlets'{, discussed in sect.~\ref{fem}}.



\section{Nodal cilia}
\label{nodal}
{Embryonic nodal cilia were first discovered in the 1990s, and a series of experimental studies confirmed that cilia motility produced a fluid flow essential for the breaking of left-right symmetry, resulting in the normal asymmetric placement of the internal organs in vertebrates. One of the most remarkable aspects of this work was the resolution of the long standing clinical question of how cilia dysfunction and \textit{situs inversus}, reversal of the internal body plan, often appear together \cite{Hirokawa09,Berdon04a}. In this section we review recent fluid mechanical models of this process, before presenting new results inspired by recent biological observations on the developmental progression of cilia placement and configuration.}

\subsection{Geometry of the embryonic mouse node}
Vertebrate development requires the establishment of three body axes. In order of appearance they are: dorsal-ventral, anterior-posterior, and left-right. Following a series of experimental studies, it is now known that cilia motion converts the already-established anterior-posterior axis information with an intrinsic chirality \cite{hilfinger2008chirality}, i.e.\ rotational direction, into left-right asymmetric flow. Cilia perform a `whirling' clockwise motion, viewed tip-to-base, with axis of rotation tilted towards the posterior. This model was first advanced by Cartwright \textit{et al.\ }\cite{Cartwright04}, with the fluid mechanics of cell surface interaction being discussed by Brokaw \cite{Brokaw05} and analysed in more detail by Smith \textit{et al.\ }\cite{Smith11,Smith07,Smith08}. Hashimoto \textit{et al.\ }\cite{Hashimoto10} established that the posterior tilt is generated because the basal body of the cilium migrates towards the posterior side of the convex cell surface.

This process takes place in the ventral node, shown in fig.~\ref{fig:mouse_node}. In the most extensively-studied species, the mouse, the node is {an approximately} triangular depression measuring $50$--$100\,\mu$m in width and $10$--$20\,\mu$m in depth \cite{Hirokawa09}, forming on the ventral side of the embryo at $7$--$9$ days post-fertilisation (dpf). The node is covered with a membrane and filled with extraembryonic fluid, which we model as Newtonian, motivating an approach based on the Stokes flow theory described in sect.~\ref{sec:fluid_dynamics}. The dorsal surface of the node is covered with a few hundred nodal pit cells that typically have one or two cilia protruding from them into the node. Each cilium is approximately $3$--$5\,\mu$m in length and $0.3\,\mu$m in diameter.

Hashimoto \textit{et al.\ }\cite{Hashimoto10} characterised morphological\\changes during the developmental process that primarily affect cilia numbers and positions, which we summarise briefly: at the \textit{late bud} stage, occurring at approximately $7.5$--$8$ dpf (for details of developmental stages, see Downs \& Davies \cite{Downs93}), cilia exhibit a distribution of tilt angle, but no overall bias. At a slightly later stage of development, \textit{early headfold}, also occurring in the period approximately $7.5$--$8$ dpf and subsequently \textit{late headfold}, at approximately $8$ dpf, cells exhibit a significant posterior bias, and a significantly greater posterior bias in the peripheral regions relative to the centre. In the later developmental stages known as \textit{1 somite} and \textit{3 somite}, occurring at approximately 8 dpf, posterior tilted cilia are extremely dominant.


During development, cilia migrate towards the posterior side of the nodal pit cell, which results in an increase of tilt angle in the posterior direction. These features will be modelled using different sets of cilia positions and parameters. {As observed by Hashimoto \textit{et al.\ }\cite{Hashimoto10} certain} mutant embryos with \textit{Dvl1, Dvl2, Dvl3} genes `knocked out' do not exhibit cilia migration and consequently do not produce a directional fluid flow. The experimental study {of Hashimoto \textit{et al.\ }} provides information on instantaneous slices of the two-dimensional flow fields using microscopic particle imaging velocimetry and how this is altered by genetic knockouts; a complementary technique to study the flow generation and the influence of cilia tilt is to formulate a mathematical model, allowing the prediction of features which are not yet available experimentally, for example three-dimensional particle tracks. We will describe briefly a computation model of Stokes flow generated by tilted rotating cilia published previously \cite{Smith11} and then apply the model to interpret the fluid mechanics of the changing developmental stages described by Hashimoto \textit{et al.}

\subsection{Modelling the stages of development in the embryonic mouse node}

We will represent the ciliated surface, {or nodal floor,} by the plane $x_3 = 0$ with cilia protruding into the region $x_3 > 0$. The $x_1$-axis will correspond to the left-right axis, with positive $x_1$ being towards the `left', and the $x_2$-axis will be the anterior-posterior axis, with negative $x_2$ being towards the posterior. In this configuration, posterior tilt corresponds to the negative $x_2$-direction.

Each cilium is tilted towards the already-established posterior direction by an angle $\theta$ and performs a conical rotation with semi-cone angle $\psi$ and angular frequency $\omega$. The centreline at arclength $s$ and time $t$ is therefore given by
{\begin{subequations}
	\begin{align}
		\xi_1(s,t) =& s\sin\psi\cos(\omega t), \\
		\xi_2(s,t) =& s(-\sin\psi\sin(\omega t)\cos\theta - \cos\psi\sin\theta), \\
		\xi_3(s,t) =& s(-\sin\psi\sin(\omega t)\sin\theta + \cos\psi\cos\theta),
	\end{align}
	\label{eq:tilted}
\end{subequations}}
with the restriction $\theta + \psi < 90\degrees$ so that each cilium does not come into contact with the ciliated surface, see fig.~\ref{fig:Cilia}. The slenderness ratio is defined as $\eta=a/L$, where $a$ is the cilium radius and $L$ is the cilium length. For nodal cilia $\eta$ is approximately $0.1$. The relative slenderness of the cilium indicates that slender body theory can be used to represent both the near- and far-field flows accurately, however it is also necessary to take into account both the no-slip condition on the plane boundary representing the ciliated surface, and the membrane that encloses the node, known as \textit{Reichert's membrane}.

\begin{figure}[ht]
	\centering
	\caption{Copyright figure available in published version. The configuration of a tilted straight rod by an angle of
	$\theta$, where $\psi$ is the semi-cone angle. Axis notation, V,
ventral; A, anterior; P, posterior; L, left and R, right.}	
\label{fig:Cilia}
\end{figure}
The no-slip condition on the plane boundary is satisfied through the Stokeslet and image system given in eq.~\eqref{eq:blakelet}. The no-slip condition on the cilium is satisfied by a Stokeslet and quadratically-weighted source dipole distribution.
In this formulation the surface of a cilium is modelled as a slender prolate ellipsoid given by,
\begin{align}
\bm{X}_\alpha(s,t) =& \bm{\xi}(s,t) \notag \\
&\hspace{-4em}+ a\sqrt{1 - \frac{(s-1/2)^2}{a^2 + 1/4}}(\bm{n}(s,t)\cos\alpha + \bm{b}(s,t)\sin\alpha).
\label{eq:ellipsoid}
\end{align}
The vectors $\bm{n}(s,t)$ and $\bm{b}(s,t)$ are normal and binormal respectively, the azimuthal angle $\alpha$ ranges from $0$ to $2\pi$, and $\bm{\xi}(s,t)$ is the straight centreline of the ellipsoid defined by eq.~\eqref{eq:tilted}. To ensure that $u_i(\bm{X}_\alpha(s_0,t)) \approx \partial_t\xi_i(s_0,t)$ uniformly, including towards the cilium ends and for all $\alpha$, we combine a centreline distribution of Stokeslets with a quadratically-weighted distribution of source dipoles \cite{Johnson80}.

The flow due to a single cilium $\hat{u}_i$ is then represented by the integral equation,
\begin{align}
\hat{u}_i(\bm{x},t) =& \int_0^1\left[B_{ij}(\bm{x};\bm{\xi}(s,t))f_j(s,t)\right. \notag \\
&\hspace{6em}\left.+ D_{ij}(\bm{x};\bm{\xi}(s,t))g_j(s,t)\right]\,\mathrm{d}s.
\label{eq:sbt}
\end{align}
The function $f_j(s,t)$ is the force per unit length on a cilium and $g_j(s,t)$ is a source dipole distribution. An image system is not required for the source dipole because the additional terms needed to satisfy the no-slip condition decay rapidly; an expression for the source dipole image system can also be derived \cite{Blake74a}. The source dipole distribution, originally derived by Chwang \& Wu \cite{Chwang75} in the context of exact solutions to Stokes flow and later found asymptotically by Johnson \cite{Johnson80}, takes the form
\begin{equation}
g_j(s,t) = -\frac{a^2s(1-s)}{\mu}f_j(s,t)\mbox{.}
\end{equation}

Due to the linearity of Stokes' equations the contribution to the velocity field from each cilium is given by the sum of slender body integrals,
\begin{align}
u_i^{\textrm{cilia}}(\bm{x},t) =& \sum_{m = 1}^{M}\int_0^L{G_{ij}(\bm{x},\bm{\xi}^{(m)}(s,t))f_j^{(m)}(s,t)\,\mathrm{d}s} \notag \\
&+ \mathcal{O}(\eta^2) \mbox{.}
\label{eq:cilia-vel}
\end{align}
{The total number of cilia is denoted by $M$}, the parameter $L$ is the length of each cilium, and the Green's function $G_{ij} = B_{ij} - a^2s(1 - s)D_{ij}/\mu$ is a combination of Stokeslet and plane boundary image system with the quadratically-weighted source dipole. The no-slip condition on the surface of each cilium is preserved provided that the cilia do not approach each other too closely, a satisfactory approximation for nodal cilia \cite{Smith07}. The centreline of the $m$th cilium is denoted $\bm{\xi}^{(m)}(s,t)$, defined by eq.~\eqref{eq:tilted} with a range of base positions. The a priori unknown force per unit length on the $m^{\textrm{th}}$ cilium is denoted $f_j^{(m)}(s,t)$.

\begin{figure}[ht!]
	\centering
		\includegraphics{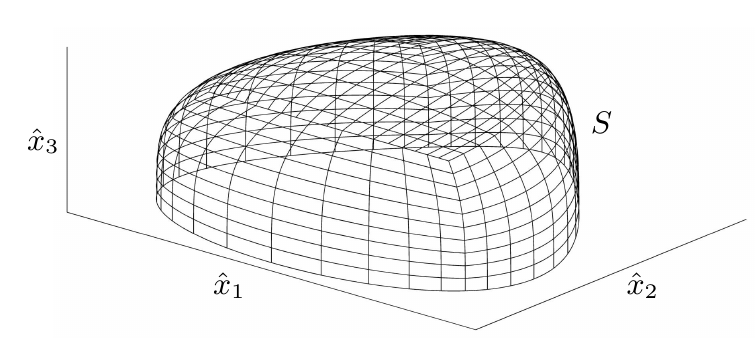}
   	\caption{A view of the mesh used to enclose the node {denoted by $S$}. The $\hat{x}_1$ axis represents the left-right axis with positive $\hat{x}_1$ being towards the left of the embryo. The $\hat{x}_2$ axis represents the anterior-posterior axis with negative $\hat{x}_2$ being towards the posterior. The $\hat{x}_3$ axis represents the dorsal-ventral axis with positive $\hat{x}_3$ being the ventral direction.}
	\label{fig:mesh}
\end{figure}

To incorporate the covering membrane of the mouse node a surface mesh of a cube is rearranged into a smooth, approximately triangular shape $S$, as described in Smith \textit{et al.\ }\cite{Smith11} and shown in fig.~\ref{fig:mesh}.
The contribution to the velocity field from the membrane can be approximated by
a single-layer boundary integral of regularised Stokeslets and plane boundary
images over the membrane surface $S$. {To obtain a regular flow
field throughout domains containing singularity distributions, Cortez \cite{Cortez01} developed the method of regularised Stokeslets. A regularised Stokeslet is defined as the exact solution to eq.~\eqref{eq:stokes} where $\bm{F}$ is given as a smoothed point-force, $\bm{F} = \bm{f}\psi_\epsilon(\bm{x} - \bm{y})$. The symbol $\psi_\epsilon(\bm{x} - \bm{y})$ denotes a cut-off function with regularisation parameter $\epsilon$. For the choice $\psi_\epsilon(\bm{x} - \bm{y}) = 15\epsilon^4/(8\pi\mu r_\epsilon^7)$, Cortez \textit{et al.\ }\cite{Cortez05} showed that the regularised Stokeslet has the form,
\begin{equation}
S_{ij}^{\epsilon}(\bm{x},\bm{y}) = \frac{1}{8\pi\mu}\left(\frac{\delta_{ij}(r^2 + 2\epsilon^2) + r_ir_j}{r_\epsilon^3}\right)\mbox{,}
\label{eq:reg_sto}
\end{equation}
where $r_\epsilon^2 = r^2 + \epsilon^2$ and the velocity due to a regularised Stokeslet in an infinite domain is then $u_i = S_{ij}^\epsilon f_j$. Ainley \textit{et al.\ }\cite{Ainley08} then derived the regularised image system Stokeslet that satisfies the no-slip boundary condition which can be written in index notation for a plane boundary at $x_3 = 0$ as
\begin{align}
      B_{ij}^{\epsilon}(\bm{x},\bm{y}) &= \frac{1}{8\pi\mu}\left(\frac{\delta_{ij}(r^2 + 2\epsilon^2) + r_ir_j}{r_{\epsilon}^3}\right. \notag \\
      &\hspace{15ex}- \frac{\delta_{ij}(R^2 + 2\epsilon^2) + R_iR_j}{R_{\epsilon}^3} \notag \\
      &\hspace{-1ex}+ 2h\Delta_{jk}\left[\frac{\partial}{\partial R_k}\left\{\frac{hR_i}{R_{\epsilon}^3} - \frac{\delta_{i3}(R^2 + 2\epsilon^2) + R_iR_3}{R_{\epsilon}^3}\right\}\right. \notag \\
      &\hspace{-1ex}\left.\left.- 4\pi h\delta_{ik}\phi_{\epsilon}(R)\vphantom{\frac{1}{1}}\right] - \frac{6h\epsilon^2}{R_{\epsilon}^5}(\delta_{i3}R_j - \delta_{ij}R_3)\vphantom{\frac{1}{1}}\right) \label{eq:reg_blakelet}
\end{align}
where $R_\epsilon^2 = R^2 + \epsilon^2$, $\phi_\epsilon(R) = 3\epsilon^2/(4\pi R_\epsilon^5)$ and $\Delta_{jk}$ has the same definition as in eq.~\eqref{eq:blakelet}. The velocity contribution from the covering membrane is then,}
\begin{equation}
u_i^{\textrm{mem}}(\bm{x},t) = \iint_S{B_{ij}^{\epsilon}(\bm{x},\bm{y})\phi_j(\bm{y},t)\,\mathrm{d}S_{\bm{y}}} + \mathcal{O}(\epsilon^2),
\label{eq:membrane-vel}
\end{equation}
with $\phi_j(\bm{y},t)$ the unknown $j^{\textrm{th}}$ component of the stress on the membrane at $\bm{y}$ and $\epsilon$ a regularisation parameter, in this case chosen to be $0.06L$ (for further details, see references \cite{Smith11,Cortez01,Cortez05,Ainley08,Smith09b}). An advantage of a regularised Stokeslet formulation in this context is that the velocity field near to, and on, the surface $S$ is regular, and can be calculated using Gauss-Legendre quadrature.

The velocity in the domain {can then be written as the sum,}
\begin{align}
u_i(\bm{x},t) = \iint_S{B_{ij}^{\epsilon}(\bm{x},\bm{y})\phi_j(\bm{y},t)\,\mathrm{d}S_{\bm{y}}} &+ \mathcal{O}(\epsilon^2) \notag \\
&\hspace{-34ex}+ \sum_{m = 1}^{M}\int_0^L{G_{ij}(\bm{x},\bm{\xi}^{(m)}(s,t))f_j^{(m)}(s,t)\,\mathrm{d}s} + \mathcal{O}(\eta^2).
\label{eq:domain-vel-init}
\end{align}
It remains to approximate the unknown stress distribution $\phi_j(\bm{y},t)$ and force per unit length distribution $f_j^{(m)}(s,t)$ for each timestep comprising a beat cycle. This is achieved by imposing the no-slip condition at discrete points on each cilium and the surface mesh, and using a boundary element constant force discretisation; for full details of the numerical implementation, see refs. \cite{Smith11,Smith09b}. {Once the stress and force per unit length distributions are known, it is then possible to calculate the fluid velocity from eq.~\eqref{eq:domain-vel-init}; knowledge of the velocity field then enables simulation of particle transport.}

\subsection{Configurations of cilia during development}
Since computational expense grows approximately with the {cube} of the number of cilia, we model the node with a smaller number of cilia than are observed experimentally. However, the underlying characteristics of the problem such as tilt direction and the geometry of the problem domain will be preserved.

{Building on the work of Smith \textit{et al.\ }\cite{Smith11}, we will focus on the stages of development where the fluid flow appears to change from vortical to directional; the stages considered are the late bud, early headfold, late headfold, 1 somite and 3 somite.} For the late bud stage $17$ cilia were distributed in the centre of the node with tilt angles in the range $-10\degrees\leq\theta\leq10\degrees$. A negative tilt angle denotes an anterior tilt and $7/17$ cilia were tilted in the anterior direction. The early headfold stage is modelled with $21$ cilia with tilt angles in the range $-10\degrees\leq\theta\leq15\degrees$ with $3/21$ cilia tilted in the anterior direction. The late headfold stage is modelled with $25$ cilia with tilt angles in the range $-5\degrees\leq\theta\leq20\degrees$ with only $1/25$ cilia tilted in the anterior direction. The 1 somite stage is modelled with $28$ cilia with tilt angles in the range $20\degrees\leq\theta\leq35\degrees$ and the 3 somite stage is modelled with $28$ cilia with tilt angles in the range $35\degrees\leq\theta\leq45\degrees$. Thus we model the increased numbers of cilia and increases in posterior tilt occurring with each advance with developmental stage, consistent with the experimental observations of approximately $150$ cilia at the late bud stage and approximately $380$ cilia at the 3 somite stage.

\subsection{Particle transport}
Results presented in this section will adopt the convention of the `left' of the node on the right of the figure, with the following set of scalings: length is normalised with respect to cilium length, which is typically $3$--$5\,\mu$m for the mouse. Time is scaled with respect to a beat period, $2\pi/\omega$, where $\omega$ is the angular frequency and velocities are then scaled according to these length and time scalings.

\begin{figure}[ht!]
	\centering
            \includegraphics{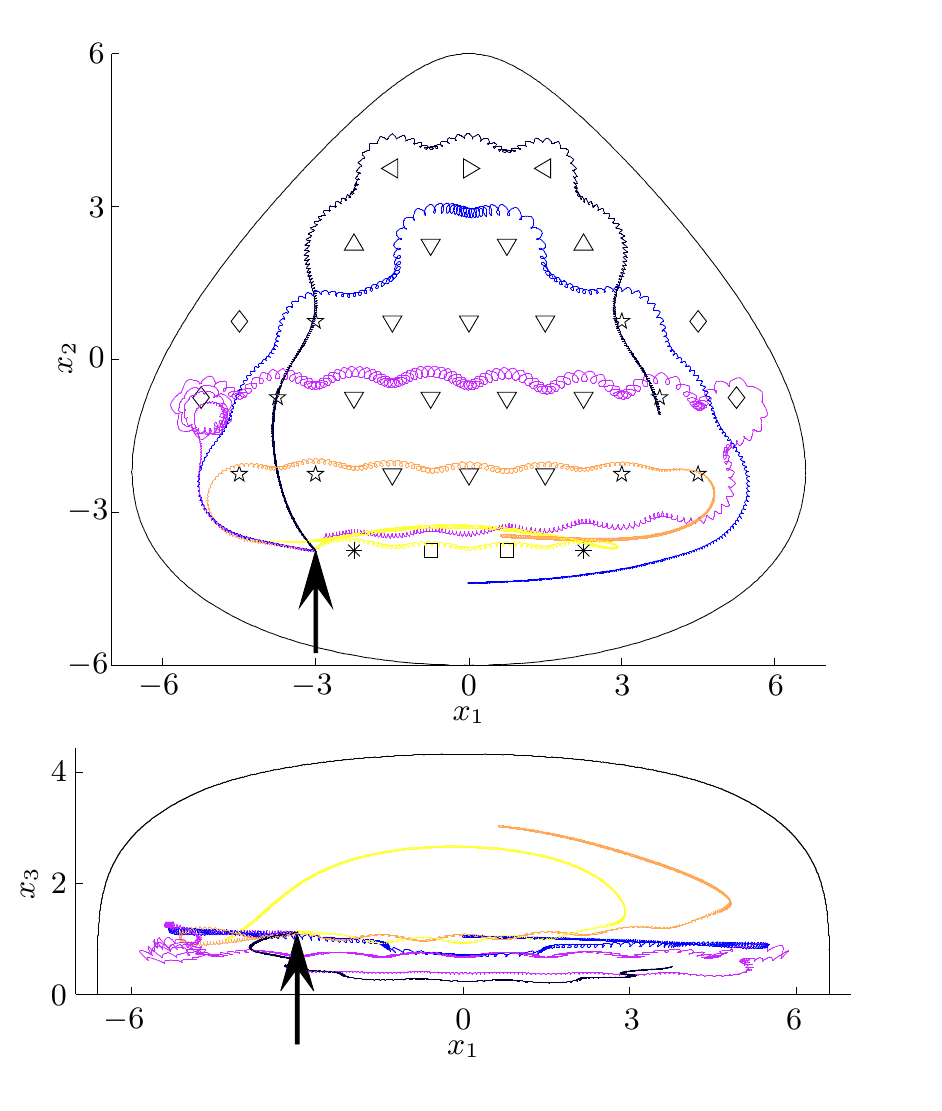}
	\caption{Particle paths for the late bud ({\bf{\color{custom1}---}}), early headfold ({\bf{\color{custom2}---}}), late headfold ({\bf{\color{custom3}---}}), 1 somite ({\bf{\color{custom4}---}}) and 3 somite ({\bf{\color{custom5}---}}) stages of development. Note that the right hand side of the figure (positive $x_1$) corresponds to the eventual `left' axis of the embryo. Cilia positions are denoted by $\triangledown$, $\triangleleft$, $\triangleright$, $\triangle$, $\medstar$, $\diamond$, $\square$, $\ast$, where $\triangledown$ are present at all stages, $\triangleleft$, late bud only, $\triangleright$, late bud, early headfold, late headfold only, $\triangle$, late bud, 1 somite only, $\medstar$, early headfold, late headfold, 1 somite, 3 somite only, $\diamond$, late headfold, 1 somite, 3 somite only, $\square$, 1 somite, 3 somite only, $\ast$, 3 somite only. The initial position for each trajectory is marked with an arrow at $x_1 = -3.00$, $x_2 = -3.75$, $x_3 = 1.10$, showing the $x_1x_2$ projection in the upper panel and the $x_1x_3$ projection in the lower panel.}
	\label{fig:sim1_2}
\end{figure}

\begin{figure}[ht!]
	\centering
            \includegraphics{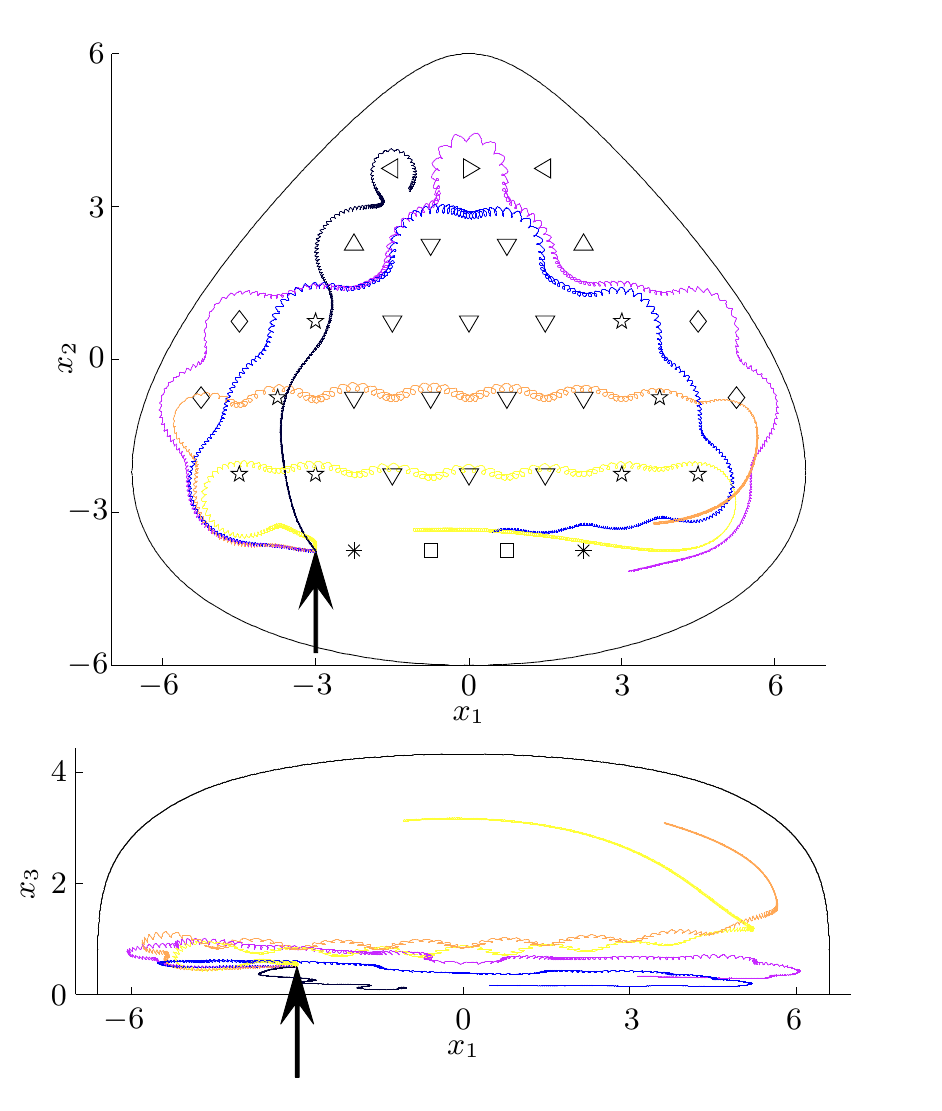}
	\caption{Particle paths for the late bud, early headfold, late headfold, 1 somite and 3 somite stages of development. Cilia positions are denoted as in fig.~\ref{fig:sim1_2}. The initial position for each trajectory is marked with an arrow at $x_1 = -3.00$, $x_2 = -3.75$, $x_3 = 0.50$, showing the $x_1x_2$ projection in the upper panel and the $x_1x_3$ projection in the lower panel.}
	\label{fig:sim2_2}
\end{figure}

\begin{figure}[ht!]
	\centering
            \includegraphics{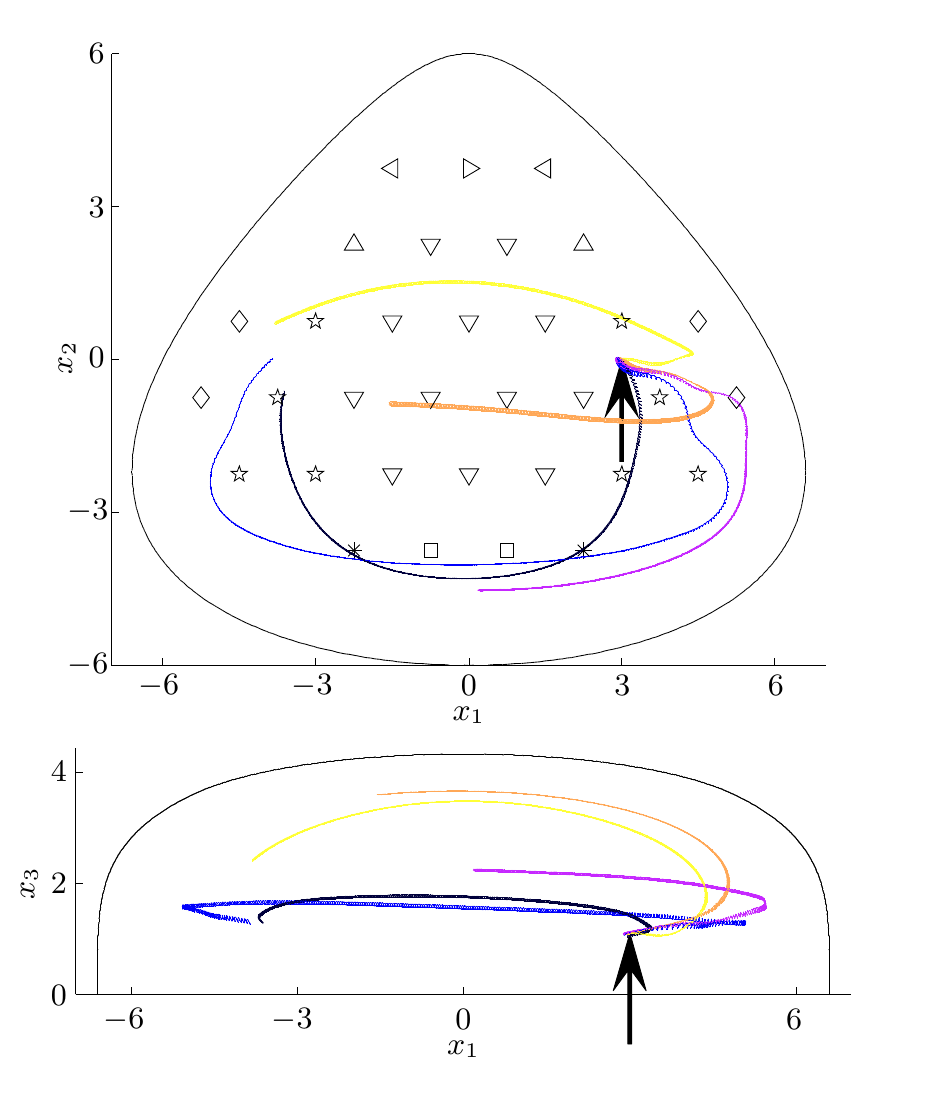}
\caption{As fig.~\ref{fig:sim2_2}, with initial particle position at $x_1 = 3.00$, $x_2 = 0.00$, $x_3 = 1.10$.}
	\label{fig:sim1_12}
\end{figure}

Figs.~\ref{fig:sim1_2}, \ref{fig:sim2_2} and \ref{fig:sim1_12} show the results of particle tracking simulations. In these simulations, one beat cycle takes $60$ timesteps and the tracking simulations were run for $20,000$ timesteps or approximately $333$ beat cycles, corresponding to $30$ seconds of cilium rotation at $10$~Hz. Initial particle positions are shown with an arrow. Particles released in the late bud and early headfold stages of development are swept around the node, in a clockwise vortex when viewed from the ventral side (figs.~\ref{fig:sim1_2}, \ref{fig:sim2_2}). This is because each cilium has a low tilt angle, thereby generating a local vortex; these local vortices combine to form a larger global vortex.

Particle paths in the late headfold stage vary with position in the node. A particle released in the right of the node at initial height $x_3 = 1.1$ is advected initially to the right and then to the left by neighbouring cilia (fig.~\ref{fig:sim1_2}). A particle with initial height $x_3 = 0.5$ (fig.~\ref{fig:sim2_2}) is advected in a vortical flow around the entire node.

All particles released in the 1 somite and 3 somite stages are advected leftward by a succession of cilia (figs.~\ref{fig:sim1_2}, \ref{fig:sim2_2}). Once the paths reach the edge of the cilia array at the left of the node they return via a rightward path close to the covering membrane. A leftward particle path is observed because all cilia are tilted towards the posterior at the 1 somite stage of development. The behaviour of particles released above the cilia tips in the left region of the node is shown in fig.~\ref{fig:sim1_12}. At all stages of development the particle path shows the expected return flow characteristics of each stage.

{The change from a global vortex to a directional flow effectively breaks the symmetry of the left-right axis. This is because particles may be moved towards the left by a succession of posteriorly tilted cilia as opposed to being carried in a global vortex around the node by untilted cilia. The mechanism for how this flow is converted to asymmetric gene expression is still under active investigation.}

\section{Swimming in non-Newtonian fluids}
\label{fem}
{For a wide class of problems, for instance internal fertilisation
in mammals, the Stokes flow equations do not give an accurate representation of
the fluid environment. In such cases, complex fluid rheological properties can
have a significant impact upon swimming speed \cite{johnson2012femlets}, and the
need for detailed study of non-Newtonian swimming has long been recognised
\cite{mills1978flat,katz1980new}. Whilst much insight has been gained into the
effects of viscoelastic rheology
\cite{pak2011,fu2009swimming,teran2010viscoelastic}, relatively less study has
been given to understanding the impact of shear-dependent viscosity on viscous
swimming.

We will model swimming in generalised Newtonian fluids for which the effective
fluid viscosity, $\mu_{\mathrm{eff}}$, is a function of $\dot{\gamma} =
\left(\frac{1}{2} \varepsilon_{ij}(\bm{u}) \varepsilon_{ij}(\bm{u})\right)^{1/2}$, the
second invariant of the strain rate tensor $\varepsilon_{ij} = \left(\partial_j
u_i + \partial_i u_j \right)$. The governing
equations for such fluids are given by
\begin{subequations}
	\label{eq:femlet_strong}
	\begin{align}
	& \nabla \cdot
	\left(\mu_{\mathrm{eff}}(\dot{\gamma})\boldsymbol{\varepsilon}(\bm{u})
	\right) - \nabla p + \bm{F} = 0, \label{eq:femlet_strong:a}\\
	& \nabla \cdot \bm{u} = 0,  \label{eq:femlet_strong:b}
	\end{align}
\end{subequations}
which are typically nonlinear, and thus established techniques involving the
superposition of fundamental flow solutions are not appropriate.

\subsection{Swimming in shear-thinning rheology}}
{
Long polymer chains in suspension tend to untangle and
align with the flow. Such suspensions are said to be `shear-thinning', since their
effective viscosity decreases with fluid shear. A swimmer in shear-thinning fluid
creates an envelope of thinned fluid around itself, which has a non-trivial
effect on its locomotion \cite{johnson2012femlets}.

The dynamics of polymer suspensions, such as cervical mucus, may be
modelled by eq. \eqref{eq:femlet_strong} with the Carreau constitutive law
\cite{carreau1968rheological}, for which
\begin{equation}
	\mu_{\mathrm{eff}}(\dot{\gamma}) = \mu_{\infty} + (\mu_0 - \mu_{\infty})
	(1 + (\lambda \dot{\gamma} )^2 )^{(n-1)/2}. \label{eq:carreau_visc}
\end{equation}
For Carreau fluids, the effective viscosity decreases monotonically between a
zero strain rate viscosity, $\mu_0$, and an infinite strain rate viscosity
$\mu_{\infty}$. {The material constant $\lambda$ is a measure of
the polymer chain relaxation time.}

Using the scalings $\bm{u} = \omega L \hat{\bm{u}}, \bm{x} = L \hat{\bm{x}}, t =
2\pi\hat{t} / \omega, \bm{f} = \mu_\infty \omega L \hat{\bm{f}}$ and $p =
\mu_\infty \omega\hat{p}$, where $\omega$ is the angular {beat frequency} and $L$ is a characteristic length, for instance the length of {the} flagellum, the dimensionless momentum equation governing the Carreau
fluid is {
\begin{equation}
	\hat{\nabla} \cdot \left[ \left( 1 + \left[ \frac{\mu_0}{\mu_\infty} - 1
	\right]\left[1 + \left(\lambda \omega
	\hat{\dot{\gamma}}\right)^2\right]^{(n-1)/2}\right)
	\hat{\bm{\varepsilon}}({\hat{\bm{u}}}) \right] - \hat{\nabla} \hat{p} +
	\hat{\bm{F}} = 0.
	\label{eq:carreau_swimming}
\end{equation}}
For fixed beat kinematics, the trajectory is thus dependent upon three
dimensionless quantities: the ratio $\mu_0/\mu_\infty$ of the zero to infinite
shear rate viscosities, the product $\lambda \omega$ of the characteristic
relaxation time of the fluid with the angular {beat} frequency,
known as the Deborah number $\mathrm{De}$, and the power-law index $n$.}

\subsection{{Modelling non-Newtonian swimming with femlets}}
{
We use the method of femlets, described in Montenegro-Johnson \textit{et al.}
\cite{johnson2012femlets}, to model viscous swimming in a generalised Newtonian
fluid.
The method of regularised Stokeslets and the method of femlets represent the
interaction of the swimmer with the fluid through a set of concentrated `blob'
forces of unknown strength and direction. While the method of regularised
Stokeslets reduces the problem to finding the coefficients in a linear
superposition of velocity solutions of known form, the method of femlets
proceeds by applying the finite element method to solve simultaneously the fluid
velocity field and the strength and direction of the forces. The use of the
finite element method removes the need for the governing equations to be linear.
Henceforth, we will continue to use dimensionless variables, but doff hats for
conciseness.

Let $D$ be a bounded domain in $\mathbb{R}^d$, where in our case $d = 2$.
We partition the domain boundary $\partial D = \partial D_{\mathrm{dir}} \cup
\partial D_{\mathrm{neu}}$ into those portions on which Dirichlet and Neumann
type boundary conditions are applied respectively. The surface of the swimmer,
$\partial D_{\mathrm{swim}} \subset \partial D_{\mathrm{dir}}$ forms a part of
the Dirichlet boundary. However, we will model the interaction of $\partial
D_{\mathrm{swim}}$ with the fluid by an immersed body force distribution. Thus for our case
$\partial D_{\mathrm{swim}}$ is not a domain boundary, but rather a manifold
containing points within the domain.

Let $H^1(D)$ be the standard Sobolev space of weakly differentiable functions
\cite{braess2007finite} defined on $D$, and
\begin{subequations}
	\begin{align}
	V_E =& \left\{\bm{w}\in (H^1(D))^d: \bm{w}_{\mid_{ \partial D_{\mathrm{dir}}}}
	= \bm{u}_{\mathrm{dir}}\right\}, \\
	V_0 =& \left\{\bm{w}\in (H^1(D))^d: \bm{w}_{\mid_{ \partial D_{\mathrm{dir}}}}
	= {\bm{0}}\right\},
	\end{align}
\end{subequations}
where $\bm{u}_{\mathrm{dir}}$ are the Dirichlet conditions imposed on $\partial
D_{\mathrm{dir}}$. Let also $Q = L^2(D)$. Multiplying
\eqref{eq:femlet_strong:a}, \eqref{eq:femlet_strong:b} by arbitrary `test' functions
$\bm{v} \in V_0, q\in Q$, respectively, yields the following integral form of
problem \eqref{eq:femlet_strong}:
\begin{subequations}
	\label{eq:femlet_weak}
	\begin{align}
	& \int_D \left\{ \nabla \cdot \left[\mu_{\mathrm{eff}}(\dot{\gamma})
	\boldsymbol{\varepsilon}(\bm{u}) \right] - \nabla p +
	\bm{F}\right\}\cdot\bm{v} \,\mathrm{d}\bm{x} = 0, \label{eq:femlet_weak:a} \\
	& \int_D q \nabla \cdot \bm{u} \,\mathrm{d}\bm{x} = 0.
	\label{eq:femlet_weak:b}
	\end{align}
\end{subequations}
Integration by parts yields an equivalent integral formulation with reduced
differentiability requirements for $\bm{u}$ and $p$. This is known as the weak (or
variational) formulation of the generalised Stokes flow problem
\eqref{eq:femlet_strong} and reads:
\begin{subequations}
	\label{eq:weakform}
	\begin{align}
	& \mbox{Find~}(\bm{u},p)\in V_E\times Q\mbox{~such that~}\forall
	(\bm{v},q) \in V_0\times Q, \nonumber \\
	& \int_D \mu_{\mathrm{eff}}(\dot{\gamma}) \bm{\varepsilon}(\bm{u}) :
	\bm{\varepsilon}(\bm{v}) \,\mathrm{d}\bm{x} - \int_D p \nabla \cdot \bm{v}
	\,\mathrm{d}\bm{x} \nonumber \\
	+ & \int_D \bm{F} \cdot \bm{v} \,\mathrm{d}\bm{x} = 0, \\
	& \int_D q \nabla \cdot \bm{u} \,\mathrm{d}\bm{x} = 0,
	\end{align}
\end{subequations}
so that $V_E, V_0$ are the velocity solution and test function spaces
respectively.
Applied on $\partial D_{\mathrm{neu}}$ is the open boundary condition first
proposed by Papanastasiou \textit{et al.} \cite{papanastasiou1992new}, given in
our case by
\begin{equation}
	\int_{\partial D_{\mathrm{neu}}}\bm{v}\cdot\bm{\sigma} \cdot \bm{n}
	\,\mathrm{d}\bm{x} = 0. \label{eq:open_boundary}
\end{equation}
Existence and uniqueness for
problem \eqref{eq:weakform} was shown by Baranger \textit{et al.}
\cite{baranger90} for both the power law and Carreau models. The finite
element discretisation used herein is discussed by Barrett \textit{et al.} \cite{barrett93}.

We consider swimmers in truncated channels of the form shown in fig.
\ref{fig:domain_plot}. The no-slip condition $\bm{u} = \bm{0}$ is applied on the
the channel walls $\partial D_{\mathrm{dir}}$, and on the boundaries where the
domain has been truncated $\partial D_{\mathrm{neu}}$ we apply the open boundary
condition \eqref{eq:open_boundary}. The results we  will present were obtained
in a channel of height 5 and length 11, with lengths normalised to a
characteristic length for the swimmer.

\begin{figure}[tbp]
	\centering
	\includegraphics{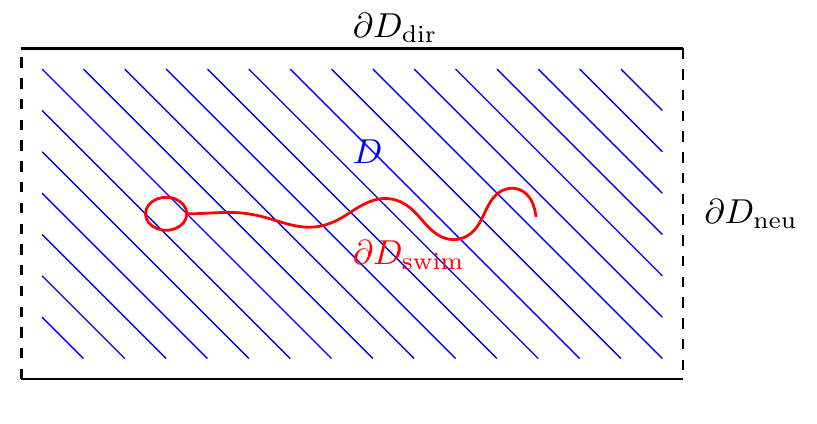}
	\caption{{An example domain $D$ containing a model human
	sperm $\partial D_{\mathrm{swim}}$ (red), showing no-slip channel walls
	$\partial D_{\mathrm{dir}}$ (solid, black) and open boundaries $\partial
	D_{\mathrm{neu}}$ (dashed, black).}}
	\label{fig:domain_plot}
\end{figure}

As the swimmer moves through the fluid, the moving boundary exerts a force
distribution on the fluid that drives the flow. We incorporate this interaction
through the unknown body force $\bm{F}$, which is governed by the motion of
the swimmer. We approximate $\bm{F}$ by a finite number of smooth immersed
forces of unknown strength and direction (femlets)
\begin{equation}
	\bm{F} = \sum_{k=1}^{N_f} g^\epsilon(\bm{x}-\bm{x}_k) \bm{f}_k,
\end{equation}	
for $N_f$ femlets of strength $\bm{f}_k$, located at $\bm{x}_k$. The cut-off
function $g^\epsilon(\bm{x}-\bm{x}_k)$ is a regularisation function similar to
that used in the method of regularised Stokeslets \cite{Cortez01}

Associated with each femlet are $K$ degrees of freedom, where $K$ is the
dimensionality of the problem domain. For example, a swimmer in two dimensions
would have the lab frame force of each femlet in the $x$ and $y$ directions,
$(f_1,f_2)$, as unknowns, resulting in $2 \times N_f$ additional scalar
variables.

To calculate the $2\times N_f$ force unknowns, we enforce $2\times N_f$
constraints in the form of Dirichlet velocity conditions $\bm{u}_s$. These are
given by the swimmer's velocity in the body frame, in which the swimmer neither
rotates nor translates, and applied at the location of each femlet. The
relationship between the body frame and the lab frame is shown in
fig.~\ref{fig:body_frame}. The body frame velocity $\bm{u}_b$ is related to the
lab frame velocity $\bm{u}$ solved for in problem \eqref{eq:weakform} by
\begin{equation}
	\bm{u}_b = \bm{u} - \bm{U} - \bm{\Omega} \times (\bm{x} - \bm{x}_0),
	\label{eq:body_lab_frame}
\end{equation}
where $\bm{U}, \boldsymbol{\Omega}$ are the {translational} and
angular velocities of the swimmer respectively and $\bm{x}_0$ is a fixed point
on the swimmer.

\setlength\fheight{6cm}
\setlength\fwidth{6cm}
\begin{figure}[tbp]
	\centering
	\includegraphics{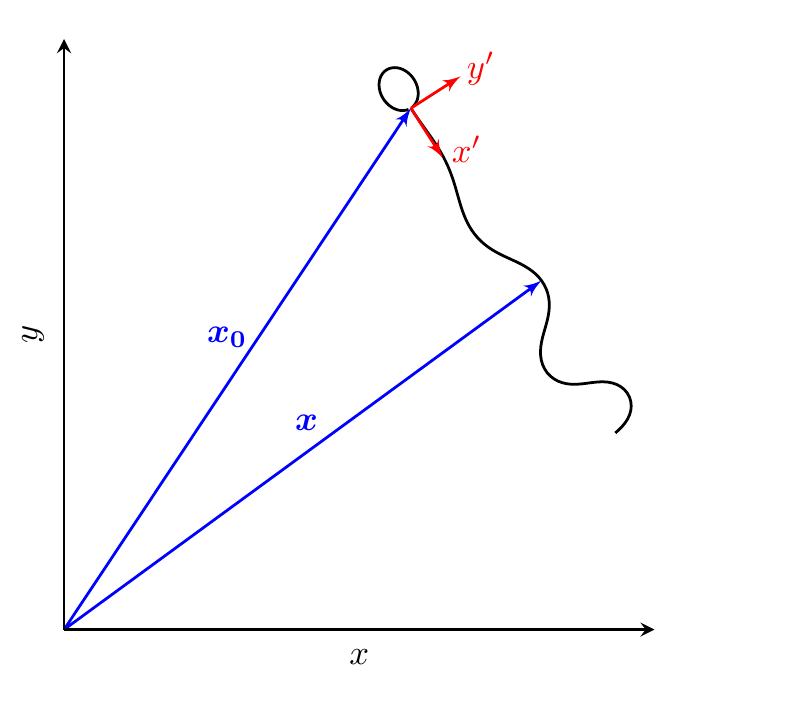}
	\caption{Schematic of the relationship between the lab frame $(x,y)$ and
	the body frame $(x^{\prime},y^{\prime})$, where $\bm{x}$ is a
	general point on the swimmer, given in the lab frame and $\bm{x_0}$
	is a reference point on the swimmer. The transformation from body frame
	velocity to lab frame velocity is given by eq.~\eqref{eq:body_lab_frame}.}
	\label{fig:body_frame}
\end{figure}

The translational and angular velocities $\bm{U}, \bm{\Omega}$ provide additional
unknowns which are closed by the conditions that zero net force and torque act
on the swimmer,
\begin{subequations}
	\begin{align}
	&\sum_{k=1}^{N_f} \bm{f}_k \int_D g^\epsilon(\bm{x}-\bm{x}_k) \,\mathrm{d}\bm{x} = 0, \\
	&\sum_{k=1}^{N_f} \bm{f}_k \times \bm{x}_k \int_D g^\epsilon(\bm{x}-\bm{x}_k) \,\mathrm{d}\bm{x} = 0.
	\end{align}
\end{subequations}
All femlets are given the same cut-off function $g^\epsilon$, and the swimming
velocity conditions are applied at the function's centroid, $\bm{x}_k$. Thus,
the force and torque conditions on the swimmer may be written
\begin{equation}
	\sum_{k=1}^{N_f}\bm{f}_k = 0, \quad \sum_{k=1}^{N_f} \bm{f}_k
	\times \bm{x}_k = 0, \label{eq:force_torque_cond}
\end{equation}
respectively. Under these conditions, problem \eqref{eq:weakform} becomes
\begin{subequations}
	\label{eq:weakform_femlets}
	\begin{align}
	& \mbox{Find~}(\bm{u},p)\in V_E\times Q\mbox{~such that~}\forall
	(\bm{v},q) \in V_0\times Q, \nonumber \\
	& \int_D \mu_{\mathrm{eff}}(\dot{\gamma}) \bm{\varepsilon}(\bm{u}) :
	\bm{\varepsilon}(\bm{v}) \,\mathrm{d}\bm{x} - \int_D p \nabla \cdot \bm{v}
	\,\mathrm{d}\bm{x} \nonumber \\
	+& \int_D \left[\sum_{k=1}^{N_f} g^\epsilon(\bm{x}-\bm{x}_k) \bm{f}_k \right] \cdot \bm{v}
	\,\mathrm{d}\bm{x} = 0, \\
	& \int_D q \nabla \cdot \bm{u} \,\mathrm{d}\bm{x} = 0, \\
	& \mbox{subject to,} \nonumber \\
	& \bm{u}(\bm{x}_k,t) = \bm{u}_s(\bm{x}_k,t) + \bm{U}(t) -
	\bm{\Omega}(t) \times (\bm{x}_k - \bm{x}_0)\mbox{,} \\
	& \sum_{k=1}^{N_f}\bm{f}_k = 0, \quad \sum_{k=1}^{N_f} \bm{f}_k
	\times \bm{x}_k = 0.
	\end{align}
\end{subequations}

Note that problem \eqref{eq:weakform_femlets} is nonlinear, due to the
dependence of $\mu_{\mathrm{eff}}$ on $\bm{u}$. We solve this nonlinear system with
the following Picard iteration: given an initial guess $(\bm{u}^0,p^0)$,
\begin{subequations}
	\label{eq:picard}
	\begin{align}
 	& \mbox{\textbf{For} }m=0,1,\ldots\mbox{ solve until convergence:} \nonumber \\
	& \mbox{Find~}(\bm{u}^{m+1},p^{m+1})\in V_E\times Q\mbox{~such
	that~} \nonumber \\
	& \forall(\bm{v},q)\in V_0\times Q, \nonumber \\
	& \int_D \mu_{\mathrm{eff}}(\dot{\gamma}^m)
	\bm{\varepsilon}(\bm{u}^{m+1}) :
	\bm{\varepsilon}(\bm{v}) \,\mathrm{d}\bm{x} - \int_D p^{m+1} \nabla \cdot \bm{v}
	\,\mathrm{d}\bm{x} \nonumber \\
	+& \int_D \left[\sum_{k=1}^{N_f}
	g^\epsilon(\bm{x}-\bm{x}_k) \bm{f}^{m+1}_k \right] \cdot \bm{v}
	\, \mathrm{d}\bm{x} = 0 \\
	& \int_D q \nabla \cdot \bm{u}^{m+1} \,\mathrm{d}\bm{x} = 0, \\
	& \mbox{subject to} \nonumber \\
	& \bm{u}(\bm{x}_k,t)^{m+1} = \bm{u}_s(\bm{x}_k,t) + \bm{U}(t)^{m+1}
	\nonumber \\
	-& \bm{\Omega}(t)^{m+1} \times (\bm{x}_k - \bm{x}_0)\mbox{,} \\
	& \sum_{k=1}^{N_f}\bm{f}_k^{m+1} = 0, \quad \sum_{k=1}^{N_f} \bm{f}_k^{m+1}
	\times \bm{x}_k = 0, \\
	& \mbox{\textbf{End.}} \nonumber
	\end{align}
\end{subequations}

At each iteration the system \eqref{eq:picard} is discretised by Taylor-Hood
P2-P1 triangular finite elements \cite{taylor1973numerical} over the domain $D$,
and the resultant linear system $\bm{M}(\dot{\gamma^{m}})\bm{z}^{m+1} = \bm{r}$
is solved. The iteration continues until
$||\bm{M}(\dot{\gamma}^{m+1})\bm{z}^{m+1} -
\bm{M}(\dot{\gamma}^{m})\bm{z}^{m+1}|| < \mathrm{tol}$, a small tolerance here
set to $\mathrm{tol} = 10^{-9}$, returning a solution vector $\bm{z}$ of the
nonlinear swimming problem. The solution, $\bm{z}$, comprises the lab frame
velocity of the fluid $\bm{u}$, the fluid pressure $p$, the force distribution
along the swimmer $\bm{f}_k$ and the swimming translational $\bm{U}$ and
rotational $\bm{\Omega}$ velocties.}

\subsection{A Najafi-Golestanian swimmer in a shear-thinning fluid}

Perhaps the simplest conceptual model of a viscous swimmer was proposed by
Najafi and Golestanian \cite{najafi2004simple}; it is highly instructive to
compare the physics of this swimmer to more detailed models of cells equipped
with cilia and flagella. The Najafi-Golestanian swimmer comprises two outer
spheres which move relative to a central sphere with a non-reciprocal motion.

The mechanism underlying the original Najafi-\\Golestanian swimmer is as
follows: one of the outer spheres will move at any given time. By force balance,
leftward relative motion of an outer sphere results in rightward motion of the
remaining spheres through the fluid, and vice versa. The distance that the
remaining spheres move in the fluid depends on the drag of the remaining two
spheres. Relative leftward motion of an outer sphere occurs while the other
spheres are far apart; relative rightward motion of an outer sphere occurs while
the other spheres are close together. Hydrodynamic interaction results in the
drag of the other spheres being reduced when they are close together. Therefore
the drag of the remaining spheres is less during relative rightward motion of
the active sphere, and so the beat cycle is slightly more effective in moving
the swimmer to the left than the right. Our variant of this swimmer follows the
suggestion {in the original paper of Najafi $\&$ Golestanian \cite{najafi2004simple} of making the sphere motion smooth in time.}

\begin{figure}[tbp]
	\begin{center}
		\includegraphics{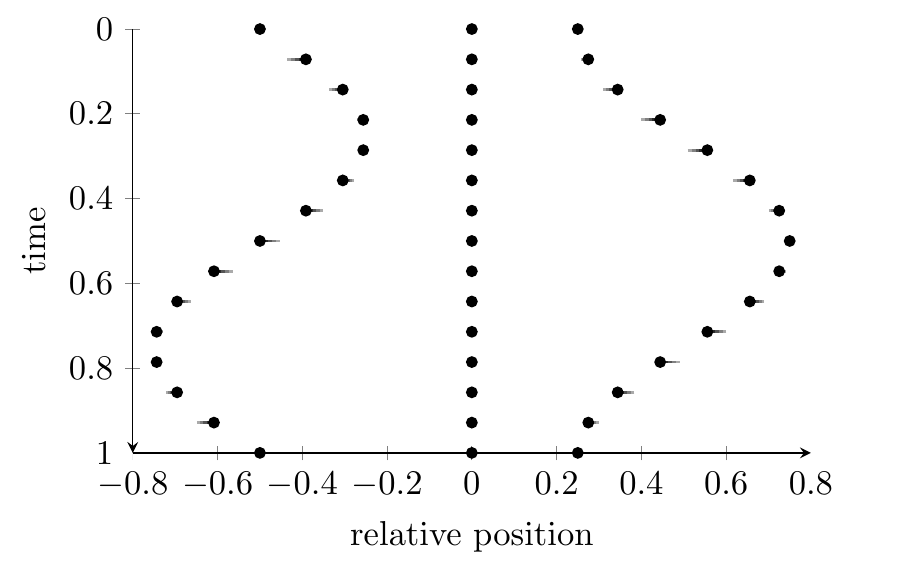}
		\caption{The position of the two outer swimming spheres, modelled
		by femlets moving relative to the central sphere for $\chi=1/4$
		over the course of a beat cycle.}	
		\label{fig:golestanian_locus}
	\end{center}
\end{figure}

{We model this swimmer by three collinear femlets}. The outer femlets move as
harmonic oscillators relative to the central femlet, and symmetry is broken by
enforcing a phase difference, $\chi$, between them
\begin{equation}
	x_1 = -d + \sin(t), \quad x_2 = 0, \quad x_3 = d + \sin(t-\chi),
\end{equation}
where $d$ is a constant displacement. The locus of this motion is shown over a
beat-cycle in fig.~\ref{fig:golestanian_locus}. Due to the symmetry of the
problem domain and beat pattern in the line $y = 0$, the swimmer will move in
the $x$-direction. Fig. \ref{fig:golestanian_position} shows the progress of
this swimmer in Stokes flow for $\chi = 1/4$, so that our example swims in the
negative $x$-direction. Here we normalise length scales such that $2d=1$.  We
will now examine the effects of {beat phase,} as well
as the Carreau rheological parameters, on {swimming speed.}
\setlength\fheight{6cm}
\setlength\fwidth{6cm}
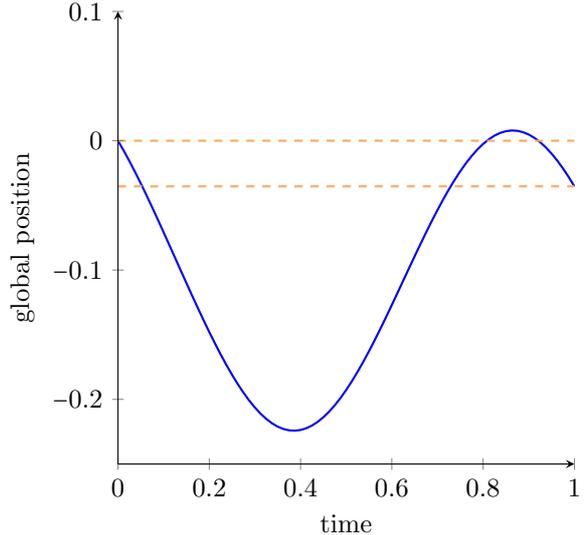
\begin{figure}[htbp]
	\begin{center}
%
%
%
%
\begin{tikzpicture}

\definecolor{mycolor1}{rgb}{0,0,0.988235294117647}
\definecolor{mycolor2}{rgb}{1,0.658039215686275,0.341960784313726}

\begin{axis}[%
width=\fwidth,
height=\fheight,
scale only axis,
xmin=0, xmax=1,
xlabel={time},
ymin=-0.25, ymax=0.1,
ylabel={global position},
axis lines=left,
axis on top]
\addplot [
color=mycolor1,
solid,thick
]
coordinates{
 (0,0)(0.0125,-0.007665317443912)(0.025,-0.0157695174304035)(0.0375,-0.0242640469080883)(0.05,-0.0331012208733963)(0.0625,-0.0422323383568549)(0.075,-0.051604831900272)(0.0875,-0.0611669439810337)(0.1,-0.0708718209364352)(0.1125,-0.0806733397822933)(0.125,-0.0905222492609475)(0.1375,-0.100371716456902)(0.15,-0.11017358124592)(0.1625,-0.119878512701361)(0.175,-0.129440316512659)(0.1875,-0.138810900722321)(0.2,-0.147942501891995)(0.2125,-0.156784059364886)(0.225,-0.16528217648047)(0.2375,-0.173388448799435)(0.25,-0.181052436311553)(0.2625,-0.188225641866715)(0.275,-0.194862314326388)(0.2875,-0.200913713762959)(0.3,-0.206340882307281)(0.3125,-0.211106897134071)(0.325,-0.215174446038827)(0.3375,-0.218516934990992)(0.35,-0.221107696555906)(0.3625,-0.222927349403983)(0.375,-0.223968889728869)(0.3875,-0.224224012480735)(0.4,-0.223688808133959)(0.4125,-0.222370337640152)(0.425,-0.220277173913085)(0.4375,-0.217424429478581)(0.45,-0.213834920484364)(0.4625,-0.209537517358909)(0.475,-0.204563198469142)(0.4875,-0.198943099106501)(0.5,-0.19271648093706)(0.5125,-0.185928517462555)(0.525,-0.178628636129145)(0.5375,-0.17086624046304)(0.55,-0.162691305856895)(0.5625,-0.154157898274673)(0.575,-0.145322955264205)(0.5875,-0.136244385731406)(0.6,-0.126983172612426)(0.6125,-0.117599861358402)(0.625,-0.108154043578194)(0.6375,-0.0987078587602363)(0.65,-0.0893234688823462)(0.6625,-0.0800618203338252)(0.675,-0.0709833344970904)(0.6875,-0.062147446559681)(0.7,-0.0536128843017462)(0.7125,-0.0454368498236399)(0.725,-0.0376736403165349)(0.7375,-0.0303734145546465)(0.75,-0.0235878669756096)(0.7625,-0.0173653825957976)(0.775,-0.0117468191267034)(0.7875,-0.00676925928981192)(0.8,-0.00246921163362638)(0.8125,0.00111934931814297)(0.825,0.00397213803702444)(0.8375,0.00606684420523157)(0.85,0.00738601214247525)(0.8625,0.0079205337623041)(0.875,0.00766583466737573)(0.8875,0.00662335854306457)(0.9,0.00480261303355202)(0.9125,0.00221248304567717)(0.925,-0.00113016308055555)(0.9375,-0.00519721347356736)(0.95,-0.00996169013581663)(0.9625,-0.0153884398007485)(0.975,-0.0214401260683227)(0.9875,-0.0280753313834901)(1,-0.0352494488281224) 
};

\addplot [
color=mycolor2,
dashed, thick
]
coordinates{
 (0,0)(1,0) 
};

\addplot [
color=mycolor2,
dashed, thick
]
coordinates{
 (0,-0.0352494488281224)(1,-0.0352494488281224) 
};

\end{axis}
\end{tikzpicture}
		\caption{The position of the central sphere of the Najafi-Golestanian
		swimmer in a channel of Stokes fluid over a beat cycle
		for $\chi = 1/4$.}	
		\label{fig:golestanian_position}
	\end{center}
\end{figure}

For the set of rheological parameters: {viscosity ratio}
$\mu_0/\mu_\infty = 2$, {Deborah number} $\lambda \omega = 1$,
and {power-law index} $n = 1/2$, we first examine the effect of
changing the phase difference $\chi$.  Fig.
\ref{fig:golestanian_progression_phase} shows the progress {of
the swimmer} as a function of $\chi$, revealing a maximum when $\chi = 1/4$ for
both Carreau and Stokes flow. This phase difference corresponds to a continuous
analogue of the beat pattern proposed in ref. \cite{najafi2004simple}. It should
be noted that for fluids with viscoelastic properties, this may not be the
optimum phase difference, due to the dependence of the fluid stress on its
deformation history. For this set of fluid parameters, the mean benefit to
progress is $3\%$, with a standard deviation of $0.5\%$. The zero progress
solutions at $\chi = 0,1/2$ correspond to the cases where the swimming spheres
are in phase and antiphase respectively. Since the maximum progress is achieved
for $\chi = 1/4$, we fix this variable for the remainder of the study.

We now wish to examine the effects of decreasing the power-law index, $n$. For
$n = 1$, the Carreau equations reduce to the Stokes flow equations. Decreasing
$n$ leads to a sharper decrease in the effective viscosity for lower strain
rates. For $\lambda \omega = 1$ and $\mu_0/\mu_\infty = 2$,
{progress of the swimmer} over a beat-cycle as a function of $n$
is displayed in fig.~\ref{fig:golestanian_progression_n}, showing that as $n$
decreases, {magnitude of progress per beat} increases.

By fixing $\lambda \omega = 1$ and $n = 1/2$, we examine the effects of varying
the viscosity ratio $\mu_0/\mu_\infty$. Fig.
\ref{fig:golestanian_progression_mu} displays the {progress of
the swimmer} over a beat-cycle as a function of this ratio, which shows that as
the infinite shear rate viscosity decreases, {magnitude of progress}
increases.

Finally, we consider the effect of altering the Deborah number $\mathrm{De}=
\lambda \omega$. This is especially important, since for a given fluid, the
power-law index and viscosity ratio are fixed {physical
parameters}. However, $\mathrm{De}$ is a function of {beat
frequency}, which for artificial swimmers may be controlled and therefore
optimised, much like the phase difference $\chi$.  For $\mu_0/\mu_\infty = 2$,
and $n = 1$, {progress} over a beat-cycle is displayed as a
function of $\mathrm{De}$ in fig.~\ref{fig:golestanian_progression_lambda},
which shows that for this particular model swimmer, optimum progress is achieved
for $\lambda \omega \approx 0.8$, so that the angular frequency of the swimmer
is approximately $4/5$ of the characteristic relaxation time of the fluid.

\setlength\fheight{6cm}
\setlength\fwidth{6cm}
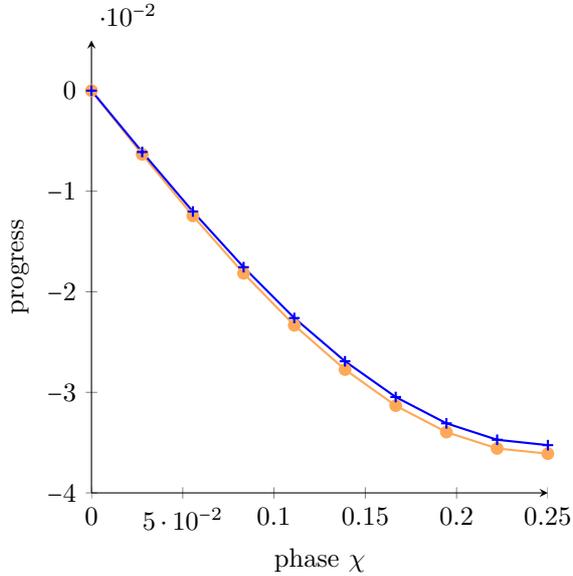
\begin{figure}[htbp]
	\begin{center}
%
%
%
%
\begin{tikzpicture}

\definecolor{mycolor1}{rgb}{1,0.658039215686275,0.341960784313726}
\definecolor{mycolor2}{rgb}{0,0,0.988235294117647}

\begin{axis}[%
width=\fwidth,
height=\fheight,
scale only axis,
xmin=0, xmax=0.25,
xlabel={phase $\chi$},
ymin=-0.04, ymax=0.005,
ylabel={progress},
axis lines=left,
axis on top]
\addplot [
color=mycolor1,
solid,thick,
mark=*,
mark options={solid,thick}
]
coordinates{
 (0,5.55111512312578e-017)(0.0277777777777778,-0.00633283466056891)(0.0555555555555556,-0.0124557342230418)(0.0833333333333333,-0.0181647331789483)(0.111111111111111,-0.0233352496675665)(0.138888888888889,-0.0277375108452842)(0.166666666666667,-0.0313256687871627)(0.194444444444444,-0.0339549857041738)(0.222222222222222,-0.0355812328511806)(0.25,-0.0360978188933247) 
};

\addplot [
color=mycolor2,
solid,thick,
mark=+,
mark options={solid,thick}
]
coordinates{
 (0,-6.93889390390723e-017)(0.0277777777777778,-0.00610110333190748)(0.0555555555555556,-0.0120193077238778)(0.0833333333333333,-0.017552516166544)(0.111111111111111,-0.0226071621965192)(0.138888888888889,-0.0269111931320199)(0.166666666666667,-0.0304667529887172)(0.194444444444444,-0.0330810146346537)(0.222222222222222,-0.0347160223107738)(0.25,-0.0352494488281224) 
};

\end{axis}
\end{tikzpicture}
		\caption{The progress of the Najafi-Golestanian swimmer over a
		single beat cycle as a function of the phase difference, $\chi$,
		for Newtonian (blue,$+$) and Carreau fluid
		(orange,\textbullet) with
		$\mu_0/\mu_\infty = 2, \lambda \omega = 1$ and $n = 1/2$.
		Negative progress denotes swimming in the negative $x$-direction.}
		\label{fig:golestanian_progression_phase}
	\end{center}
\end{figure}

\begin{figure}[htbp]
	\begin{center}
%
%
%
%
\begin{tikzpicture}

\definecolor{mycolor1}{rgb}{0,0,0.988235294117647}

\begin{axis}[%
width=\fwidth,
height=\fheight,
scale only axis,
xmin=0.2, xmax=1,
xlabel={power-law index $n$},
ymin=-0.0364, ymax=-0.0352,
ylabel={progress},
axis lines=left,
axis on top]
\addplot [
color=mycolor1,
solid,thick,
mark=*,
mark options={solid,thick}
]
coordinates{
 (1,-0.0352494488281224)(0.9,-0.0354493763376508)(0.8,-0.0356379723576516)(0.7,-0.035811317023564)(0.6,-0.035965683640122)(0.5,-0.0360978188933247)(0.4,-0.0362051978351668)(0.3,-0.0362862102150689)(0.2,-0.0363402474543442) 
};

\end{axis}
\end{tikzpicture}
		\caption{The progress of the Najafi-Golestanian
		swimmer over a single beat cycle in Carreau fluid as a function
		of the power-law index, $n$, for $\lambda \omega = 1$ and
		$\mu_0/\mu_\infty = 2$. Newtonian flow corresponds to the value
		$n = 1$. Negative progress denotes swimming in the negative
		$x$-direction.}
		\label{fig:golestanian_progression_n}
	\end{center}
\end{figure}
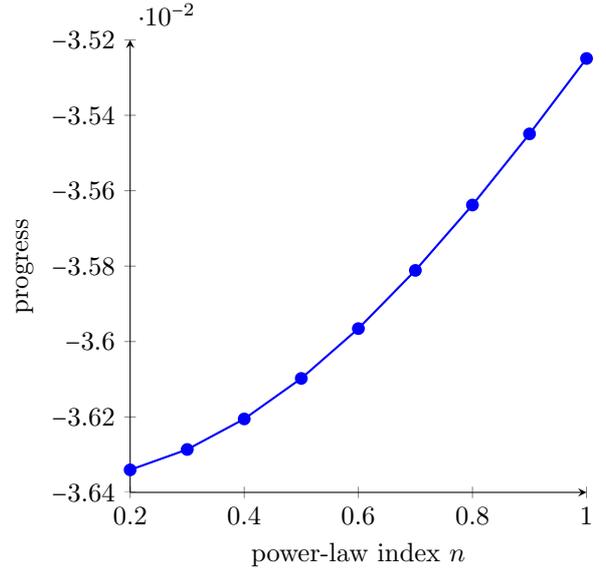

\begin{figure}[htbp]
	\begin{center}
%
%
%
%
\begin{tikzpicture}

\definecolor{mycolor1}{rgb}{0,0,0.988235294117647}

\begin{axis}[%
width=\fwidth,
height=\fheight,
scale only axis,
xmin=1, xmax=5,
xlabel={viscosity ratio $\mu_0 / \mu_\infty$},
ymin=-0.0375, ymax=-0.035,
ylabel={progress},
axis lines=left,
axis on top]
\addplot [
color=mycolor1,
solid,thick,
mark=*,
mark options={solid,thick}
]
coordinates{
 (1,-0.0352494488281224)(1.11111111111111,-0.0353809890244697)(1.25,-0.03552797465272)(1.42857142857143,-0.0356935182988072)(1.66666666666667,-0.0358816708759641)(2,-0.0360978188933247)(2.5,-0.0363493123768582)(3.33333333333333,-0.0366465066487192)(5,-0.037004602767106) 
};

\end{axis}
\end{tikzpicture}
		\caption{The progress of the Najafi-Golestanian
		swimmer over a single beat cycle in Carreau fluid as a function
		of the viscosity ratio, $\mu_0/\mu_\infty$, for $\lambda \omega
		= 1$ and $n = 1/2$, shown for equispaced values of $\mu_\infty$.
		Newtonian flow corresponds to the value $\mu_0/\mu_\infty = 1$.
		Negative progress denotes swimming in the negative $x$-direction.}
		\label{fig:golestanian_progression_mu}
	\end{center}
\end{figure}
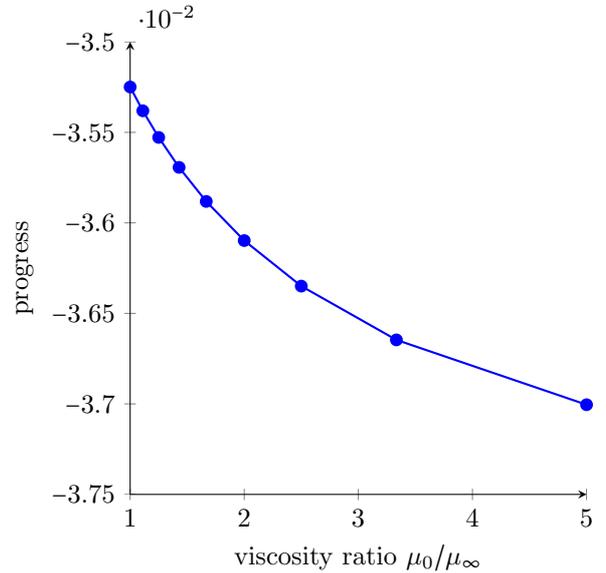

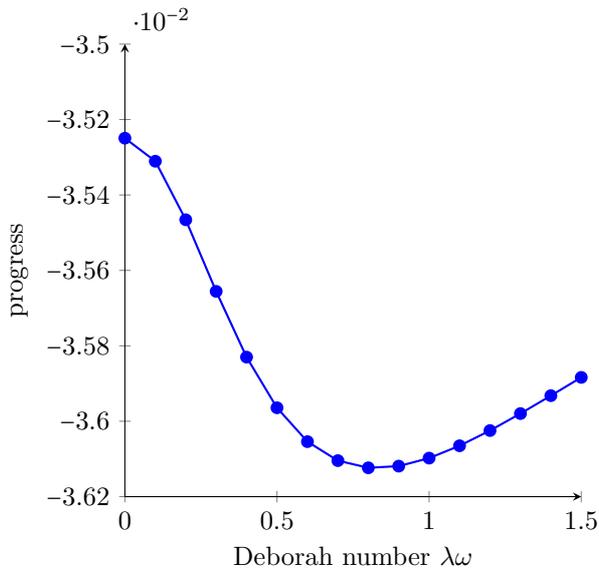
\begin{figure}[htbp]
	\begin{center}
%
%
%
%
\begin{tikzpicture}

\definecolor{mycolor1}{rgb}{0,0,0.988235294117647}

\begin{axis}[%
width=\fwidth,
height=\fheight,
scale only axis,
xmin=0, xmax=1.5,
xlabel={Deborah number $\lambda \omega$},
ymin=-0.0362, ymax=-0.035,
ylabel={progress},
axis lines=left,
axis on top]
\addplot [
color=mycolor1,
solid,thick,
mark=*,
mark options={solid,thick}
]
coordinates{
 (0,-0.0352494488281224)(0.1,-0.0353105317587195)(0.2,-0.0354657924401103)(0.3,-0.0356556768306049)(0.4,-0.0358299993338212)(0.5,-0.0359643271850187)(0.6,-0.0360541099705976)(0.7,-0.0361044673683982)(0.8,-0.0361234744272049)(0.9,-0.0361190391658242)(1,-0.0360978188933247)(1.1,-0.0360650458101982)(1.2,-0.0360246857886559)(1.3,-0.0359796808185649)(1.4,-0.0359321808416961)(1.5,-0.0358837367228409) 
};

\end{axis}
\end{tikzpicture}
		\caption{The progress of the Najafi-Golestanian
		swimmer over a single beat cycle in Carreau fluid as a function
		of the Deborah number, $\lambda \omega$, for $\mu_0/\mu_\infty =
		2$ and $n = 1/2$. Newtonian flow corresponds to the value
		$\lambda \omega = 0$. Negative progress denotes swimming in the
		negative $x$-direction.}
		\label{fig:golestanian_progression_lambda}
	\end{center}
\end{figure}

\subsection{{A two-dimensional sperm in shear-thinning and
thickening fluid}}

\begin{figure}[htbp]
	\centering
		\includegraphics{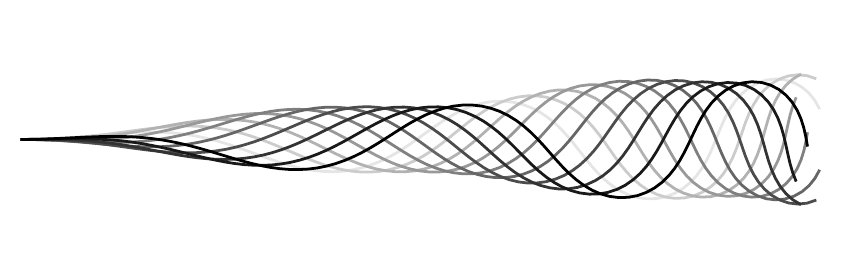}
		\caption{A model flagella waveform, generated by the shear
		angle parameterisation \eqref{shear_wave_high_visc} with
		amplitude $C = 0.9\pi/2$ and wavenumber $k = 2.5\pi$, typical of
		what is observed experimentally in high viscosity media.}
		\label{fig:waveform}
\end{figure}

For sperm in high-viscosity fluids, such as human mucus, swimming is typified by
planar flagella beating that grows in peak curvature towards the distal portion
of the tail \cite{smith2009bend}. To model this waveform, we prescribe a
body-frame tangent angle of the form
\begin{equation}
	\psi (s,t) = C s \cos(ks - t), \label{shear_wave_high_visc}
\end{equation}
for $s$ the arclength along the flagellum and $t$ time. Such a parameterisation
makes sense in the context of considering a bending wave propagating down the
flagellum, steepening towards the distal end as the stiffness of the flagellum
decreases. A representative waveform produced by the shear angle parametrisation
given by eq. \eqref{shear_wave_high_visc} is shown in
fig.~\ref{fig:waveform}.

Integrating the tangent vector along the flagellum gives the flagellar centreline
{\begin{subequations}
	\begin{align}
	x_c(s,t) =& x_0 + \int_0^s{\cos(\psi(s^{\prime},t))\,\mathrm{d}s^{\prime}}, \\
	y_c(s,t) =& y_0 + \int_0^s{\sin(\psi(s^{\prime},t))\,\mathrm{d}s^{\prime}},
	\end{align}
\end{subequations} \label{eq:planar-beat-param}}
with corresponding centreline velocity $\bm{\dot{x}}$
{\begin{subequations}
	\begin{align}
	\dot{x}_c(s,t) =& \int_0^s{-\sin(\psi)\dot{\psi}\,\mathrm{d}s^{\prime}}, \\
	\dot{y}_c(s,t) =& \int_0^s{\cos(\psi)\dot{\psi}\,\mathrm{d}s^{\prime}}.
	\end{align}
\end{subequations} \label{eq:planar-beat-vel}}

This parameterisation of the flagellum is given in {the body frame}
\cite{Higdon79}, in which the cell body neither rotates nor translates. The
translational and rotational velocities that arise from this instantaneous
configuration are then used to update the {position of the
swimmer} in the lab frame, so that the cell swims through the fluid domain.

{The trajectory of the swimmer is calculated by integrating the
swimming velocity $\bm{U}, \bm{\Omega}$ with the Adams-Bashforth multistep method
\cite{bashforth1883attempt,iserles2009first}, which interpolates an $n$th order
polynomial through the current and previous $n-1$ values of translational and
angular velocity to give the position. For a two-dimensional sperm-like swimmer
in a channel of Newtonian fluid, with less than a $0.3\%$ change in the position of the
swimmer after a single beat cycle between 25 steps per beat with the second
order scheme and 80 steps with the third order scheme. As such, we use the second
order scheme with 25 steps per beat.} We normalise length scales to the
flagellum length, so that one length unit corresponds to $55\,\mu\mathrm{m}$,
and time scales to the period of the beat. Thus, modelling a tail beating at
$10\,\mathrm{Hz}$ would mean that one time unit corresponds to
$1/10\,\mathrm{s}$.

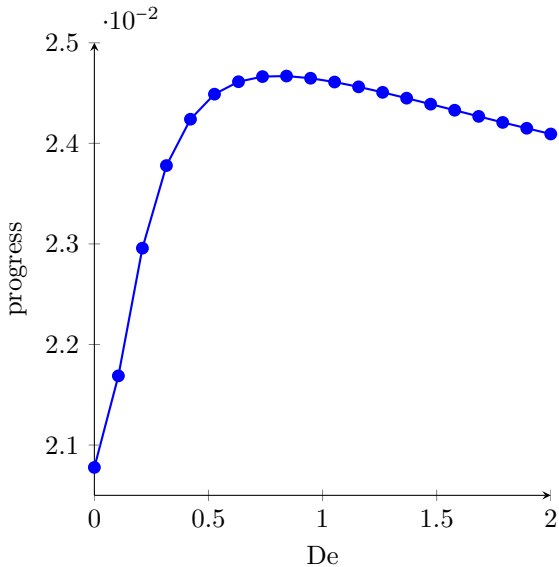
\begin{figure}[htbp]
	\begin{center}
%
%
%
%
\begin{tikzpicture}

\begin{axis}[%
width=\fwidth,
height=\fheight,
scale only axis,
xmin=0, xmax=2,
xlabel={$\mathrm{De}$},
ymin=0.0205, ymax=0.025,
ylabel={progress},
axis lines=left,
axis on top]
\addplot [
color=blue,
solid, thick,
mark=*,
mark options={solid}
]
coordinates{
 (0,0.0207776028115468)(0.105263157894737,0.0216877951382058)(0.210526315789474,0.0229577904750458)(0.315789473684211,0.0237792010506921)(0.421052631578947,0.0242399687004099)(0.526315789473684,0.0244891902476323)(0.631578947368421,0.024612913894114)(0.736842105263158,0.0246638879028965)(0.842105263157895,0.0246694659215642)(0.947368421052632,0.0246475637835239)(1.05263157894737,0.0246098880608141)(1.15789473684211,0.0245620213068944)(1.26315789473684,0.0245072782256567)(1.36842105263158,0.0244494013636922)(1.47368421052632,0.0243898221091559)(1.57894736842105,0.0243294609067899)(1.68421052631579,0.0242682078095109)(1.78947368421053,0.0242082635989111)(1.89473684210526,0.0241502231015734)(2,0.0240931952393782) 
};

\end{axis}
\end{tikzpicture}
		\caption{The unsigned total distance travelled by our two-dimensional sperm over a
		single beat cycle in Carreau fluid as a function of the Deborah
		number, $\lambda \omega$, for $\mu_0/\mu_\infty = 2$ and $n =
		1/2$. Newtonian flow corresponds to the value $\lambda \omega =
		0$.}
		\label{fig:sperm_deborah}
	\end{center}
\end{figure}

Fig. \ref{fig:sperm_deborah} shows the sperm's progress as a function of
Deborah number for $\mu_0/\mu_\infty = 2$ and $n = 1/2$. Note the qualitative
similarity to fig. \ref{fig:golestanian_progression_lambda}, with overall
progression initially increasing with $\mathrm{De}$ to a maximum value around
$\mathrm{De} = 0.8$, then decreasing at a slower rate. Since the
Najafi-Golestanian swimmer comprises three forces in balance, its representation
by a stokes quadrupole might shed insight into this link, motivated by the
observations in fig.  \ref{fig:sperm_singularity}. {These
results also draw an interesting parallel with the analysis of Teran \textit{et
al.} \cite{teran2010viscoelastic} who used the immersed boundary method to show
that the progression of a waving filament may be enhanced in a viscoelastic
Oldroyd B fluid at Deborah numbers close to 1.}

{If the viscosity ratio, $\mu_0/\mu_\infty$, is less than $1$,
then the effective viscosity \eqref{eq:carreau_visc} of a Carreau fluid
increases with shear rate. For such a model, the relaxation time $\lambda$ no
longer has a physical interpretation in terms of polymer physics, but the
Carreau law may still be used as a regularisation of a shear-thickening
power-law fluid. Examples of shear-thickening fluids are custard and a mixture
of cornstarch with water known colloquially as oobleck.}

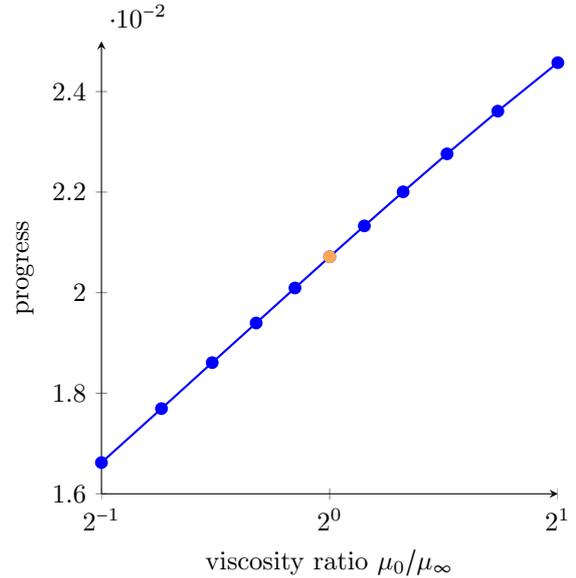
\begin{figure}[htbp]
	\begin{center}
%
%
%
%
\begin{tikzpicture}

\definecolor{mycolor1}{rgb}{0,0,0.988235294117647}
\definecolor{mycolor2}{rgb}{1,0.658039215686275,0.341960784313726}

\begin{semilogxaxis}[%
width=\fwidth,
height=\fheight,
scale only axis,
xmin=0.5, xmax=2,
xtick={0.5,1,2},
xticklabels={$2^{-1}$,$2^{0}$,$2^{1}$},
xminorticks=true,
xlabel={viscosity ratio $\mu_0/\mu_\infty$},
ymin=0.016, ymax=0.025,
ylabel={progress},
axis lines=left,
axis on top]
\addplot [
color=mycolor1,
solid,thick,
mark=*,
mark options={solid,thick}
]
coordinates{
 (0.5,0.0166212801438684)(0.6,0.0176948012285629)(0.7,0.0186091130361597)(0.8,0.0193972533939954)(0.9,0.0200938673286605)(1,0.0207137930408429)(1.11111111111111,0.0213284116773994)(1.25,0.0220059448338814)(1.42857142857143,0.0227611996436122)(1.66666666666667,0.0236108241594956)(2,0.0245746320866125) 
};

\addplot [
color=mycolor2,
only marks,
mark=*,
mark options={solid,thick}
]
coordinates{
 (1,0.0207137930408429) 
};

\end{semilogxaxis}
\end{tikzpicture}
		\caption{{The progress of the two-dimensional
		sperm-like swimmer over a single beat cycle in Carreau fluid as
		a function of the viscosity ratio, $\mu_0/\mu_\infty$, for
		$\lambda \omega = 1$ and $n = 1/2$. Stokes flow,
		$\mu_0/\mu_\infty = 1$ is marked in orange, with the
		shear-thickening r\'{e}gime for $\mu_0/\mu_\infty < 1$ and the
		shear-thinning r\'{e}gime for $\mu_0/\mu_\infty > 1$. The horizontal axis
		is displayed on a logarithmic scale	$\mu_0/\mu_\infty$.}}	
		\label{fig:sperm_progress_mu}
	\end{center}
\end{figure}

{Fig.~\ref{fig:sperm_progress_mu} shows the progress of the
sperm over a single beat-cycle as a function of the viscosity ratio
$\mu_0/\mu_\infty$. Stokes flow, corresponding to $\mu_0/\mu_\infty = 1$, is
marked in orange, and the thickening and thinning r\'{e}gimes lie to its left
and right respectively. We see that whilst shear-thinning aids progress,
shear-thickening inhibits it. Furthermore, fig.~\ref{fig:sperm_progress_mu}
shows an almost perfect linear relationship between progress and the logarithm
of $\mu_0/\mu_\infty$.}

{For shear-thinning fluids, there is a gradient of thick to thin
fluid from the proximal to distal portions of the flagellum. For shear-thickening
fluids, the gradient runs from thin to thick. Thus, the disadvantage to cell
progress in a shear-thickening fluid is consistent with the hypothesis of
Montenegro-Johnson \textit{et al.} \cite{johnson2012femlets} that differential viscosity
between the distal and proximal portions of the flagellum is responsible for
changes in propulsive efficiency.}

\section{Conclusions}
\label{conc}
\subsection{Fundamental physics of cilia and flagella driven flow}
The linearity of the Stokes flow equations makes the method of singular solutions possible. This approach provides significant insight into the basic nature of the physics of propulsion and swimming. Microscopic swimmers, for which gravitational and inertial forces are negligible, are subject to zero total force and torque, entailing that the far-field of a flagellated swimmer is stresslet in nature, the flow velocity having inverse square decay. However, within a distance of a few cell radii or flagella lengths, the flow field can be markedly different from a stresslet, being closer to a stokes quadrupole for sperm, and a stresslet-source dipole combination for single celled algae. These findings are confirmed both experimentally and through more detailed computational models. The action of gravity on larger swimmers results in a more slowly-decaying $\mathcal{O}(1/r)$ Stokeslet flow.

The lack of explicit time dependence in the Stokes flow equations entails that time-irreversible motion is essential for successful swimming and pumping. Cilia and flagella achieve this through a number of mechanisms. The anisotropy of the Stokeslet results in the anisotropic drag law for slender bodies, a property which underlies the propulsive effect of travelling bending waves. Finally, wall interaction, which can be understood by the method of images, is essential to the function of cilia. Image systems convert both forces and torques to higher order singularities in the far-field, a mechanism which explains how rotating embryonic nodal cilia create a right-to-left flow.

\subsection{Cilia in embryonic development}
The hybrid computational technique combining slender body theory with the regularised Stokeslet boundary element method allows prediction of both generic features of cilia-driven flow in the nodal cavity, as shown previously \cite{Smith11}. Our simulations show the effect of developmental changes in cilia orientation and position, using experimental data on cilia numbers and orientation \cite{Hashimoto10}. In early stages of development (late bud, early headfold), the flow is essentially vortical, whereas in later stages when the majority of cilia are tilted posteriorly (late headfold, 1 somite and 3 somite); {an asymmetric directional flow is established that breaks the symmetry of the left-right axis.}

Qualitative agreement with particle imaging velocimetry observations \cite{Hashimoto10} suggest that this approach will be useful for analysing this and related systems in more detail, providing further physical insight into the coupling of nodal flow to subsequent morphogenesis \cite{Hirokawa09,Cartwright09}.

\subsection{Swimming in non-Newtonian fluids}

The Najafi-Golestanian swimmer is able to progress because of hydrodynamic
interaction of spheres coupled with time-irreversible motion, whereas sperm
propel through propagating bending waves of a slender flagellum; however the
effect of shear-thinning rheology on both cells is qualitatively similar. Both
{models} swim faster in a shear-thinning fluid, with an optimal
value for Deborah number between 0.5 and 1, however sperm enjoy a greater
propulsive advantage relative to the Newtonian state compared with the
three-sphere swimmer, their increase being up to around {20}\%
compared with around 3\% {in the region of parameter space
investigated.} {For sperm, shear-thickening fluid rheology
led to a decrease in cell progress. It was found that cell progress had an
almost exact logarithmic dependence on the viscosity ratio $\mu_0/\mu_\infty$.
It should be noted, however, that the fluid strain-rate of a Carreau fluid is
independent of its recent deformation history, and so to fully capture the
behaviour of human mucus an extended model is desirable.}

{\subsection{Summary}}
The examples in this paper illustrate aspects of the broad range of very low Reynolds number flow. The embryonic nodal flow is typically considered Newtonian, and entails the wall effect converting rotation to directional flow, which in turn converts chiral to lateral information. The swimming cell problems illustrate the importance of more complex non-Newtonian effects typical of biological fluids such as blood and mucus, and the qualitative insights that can be gained from conceptual models.

Reproduction and development continue to be inspirational topics in the biological fluid mechanics of active matter. Future work will focus on{, among other topics,} the nonlinear interaction of multiple ciliated and flagellated cells with biological fluids and with each other. We anticipate that the techniques described will help to form a basis for these future investigations.

\subsection*{Acknowledgements}

TDMJ and AAS acknowledge Engineering and Physical Sciences Research
Council PhD studentships. DJS is funded by a Birmingham Science City Research
Alliance Fellowship. Computations in sect.~3 were performed using the University of Birmingham BlueBEAR HPC service, which
was purchased through HEFCE SRIF-3 funds.

\bibliographystyle{unsrt}
\bibliography{myrefs}

\end{document}